Protocol

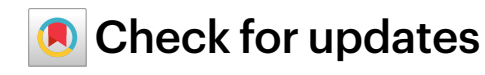

# A comprehensive framework for statistical testing of brain dynamics

Nick Y. Larsen [1], Laura B. Paulsen [1,2], Christine Ahrends [1,3,4], Anderson M. Winkler [5] & Diego Vidaurre [1,6]

## Abstract

Neural activity data can be associated with behavioral and physiological variables by analyzing their changes in the temporal domain. However, such relationships are often difficult to quantify and test, requiring advanced computational modeling approaches. Here, we provide a protocol for the statistical analysis of brain dynamics and for testing their associations with behavioral, physiological and other non-imaging variables. The protocol is based on an open-source Python package built on a generalization of the hidden Markov model (HMM)—the Gaussian-linear HMM—and supports multiple experimental modalities, including task-based and resting-state studies, often used to explore a wide range of questions in neuroscience and mental health. Our toolbox is available as both a Python library and a graphical interface, so it can be used by researchers with or without programming experience. Statistical inference is performed by using permutation-based methods and structured Monte Carlo resampling, and the framework can easily handle confounding variables, multiple testing corrections and hierarchical relationships within the data, among other features. The package includes tools developed to facilitate the intuitive visualization of statistical results, along with comprehensive documentation and step-by-step tutorials for data interpretation. Overall, the protocol covers the full workflow for the statistical analysis of functional neural data and their temporal dynamics.

[1]Center of Functionally Integrative Neuroscience, Department of Clinical Medicine, Aarhus University, Aarhus, Denmark. [2]School of Communication and Culture, Department of Linguistics, Cognitive Science and Semiotics, Aarhus University, Aarhus, Denmark. [3]Oxford Centre for Integrative Neuroimaging, Nuffield Department of Clinical Neurosciences, University of Oxford, Oxford, UK. [4]Linacre College, University of Oxford, Oxford, UK. [5]Division of Human Genetics, School of Medicine, The University of Texas Rio Grande Valley, Brownsville, TX, USA. [6]Oxford Centre for Human Brain Analysis, Psychiatry Department, University of Oxford, Oxford, UK. e-mail: nylarsen@cfin.au.dk

## Key points

- Different variants of the hidden Markov model can be used to characterize latent states in brain activity and their temporal dynamics recorded from various modalities including functional MRI, magnetoencephalography, electroencephalography, electrocorticography and local field potentials.
- This protocol presents methods for statistical inference on the relation between brain dynamics and different types of behavior.

## Key references

## Introduction

Understanding the associations between brain activity and behavior represents one of the main goals of neuroscience research[1,2], on the assumption that characterizing these brain–behavior relationships will advance our ability to manage patients with mental health and neurological disorders[3–5]. To quantify such associations, researchers typically rely on prediction techniques, statistical testing or a combination of the two. Prediction methods focus on out-of-sample accuracy, assessing how well a model generalizes to new data, whereas explanatory approaches emphasize testing formal hypotheses and identifying statistically reliable associations between variables[6].

Here, we provide easy-to-use routines for statistical testing of the relation between brain dynamics and behavioral or physiological variables and the associated Python package to run the code. The protocol builds on the Gaussian-Linear Hidden Markov Model (GLHMM) Python package[7], which implements multiple types of HMMs into a single framework for existing and new models. Using an HMM-based characterization of the data, the presented statistical framework supports a wide range of experimental designs commonly used in neuroscience, including the resting state. To support broader accessibility, we also include a graphical user interface (GUI) that enables users to run analyses without the need for writing code. The protocol covers model fitting through to result presentation, with implementation details and worked examples across modalities such as functional MRI (fMRI) and magnetoencephalography (MEG), as well as different experimental designs.

We define four families of statistical tests that address a wide range of relevant scientific questions. These are as follows: (1) across-subjects tests, which assess the associations between individual traits and brain activity across subjects; (2) across-trials tests, which compare brain activity across trials under different experimental conditions; (3) across-sessions-within-subject tests, which evaluate long-term changes in brain dynamics across multiple sessions for one subject; and (4) across-state-visits tests, which examine associations between brain time series and one or more simultaneously measured variables, such as physiological measurements.

Unlike existing frameworks for statistical inference that primarily target time-averaged or non-temporal data, this approach has a strong focus on the temporal dimension of brain activity (i.e., on brain dynamics), although it can also handle more conventional tests. Although these tests are presented with a focus on neuroscience, they are readily generalizable and can be adapted to other fields such as economics and ecology. These tools are well documented and easily generalizable to other types of data besides neuroscience. This makes the toolbox suitable for any domain that involves the statistical testing of relationships between dynamic system properties (e.g., sequential or temporal data) and a set of external variables.

### Development of the protocol

#### Estimating a model brain dynamics from time series data

We developed a framework to analyze the relationship between brain dynamics and behavior at various temporal scales through statistical testing. Brain dynamics are first characterized by using the GLHMM, a generalization of the HMM, before proceeding to the statistical testing, which forms the main focus of this paper. The HMM characterizes brain activity by using a finite set of latent states and their temporal dynamics (i.e., when they occur and the transitions between them). The GLHMM extends the standard Gaussian-state HMM by allowing multiple types of state models based on different configurations of the regression model. Leveraging this flexibility, it can be used on different brain activity modalities, including fMRI[8–12], MEG[13–16], electroencephalography (EEG)[13,17], electrocorticography (ECoG)[7] and local field potentials (LFPs)[18,19]. Specifically, the GLHMM is based on a Bayesian regression model to capture the relationship between two time series: $X$ (independent variable) and $Y$ (dependent variable). The observations are modeled as:

$$Y_t | s_t = k \sim N(\mu_k + X_t \beta_k, \Sigma_k)$$

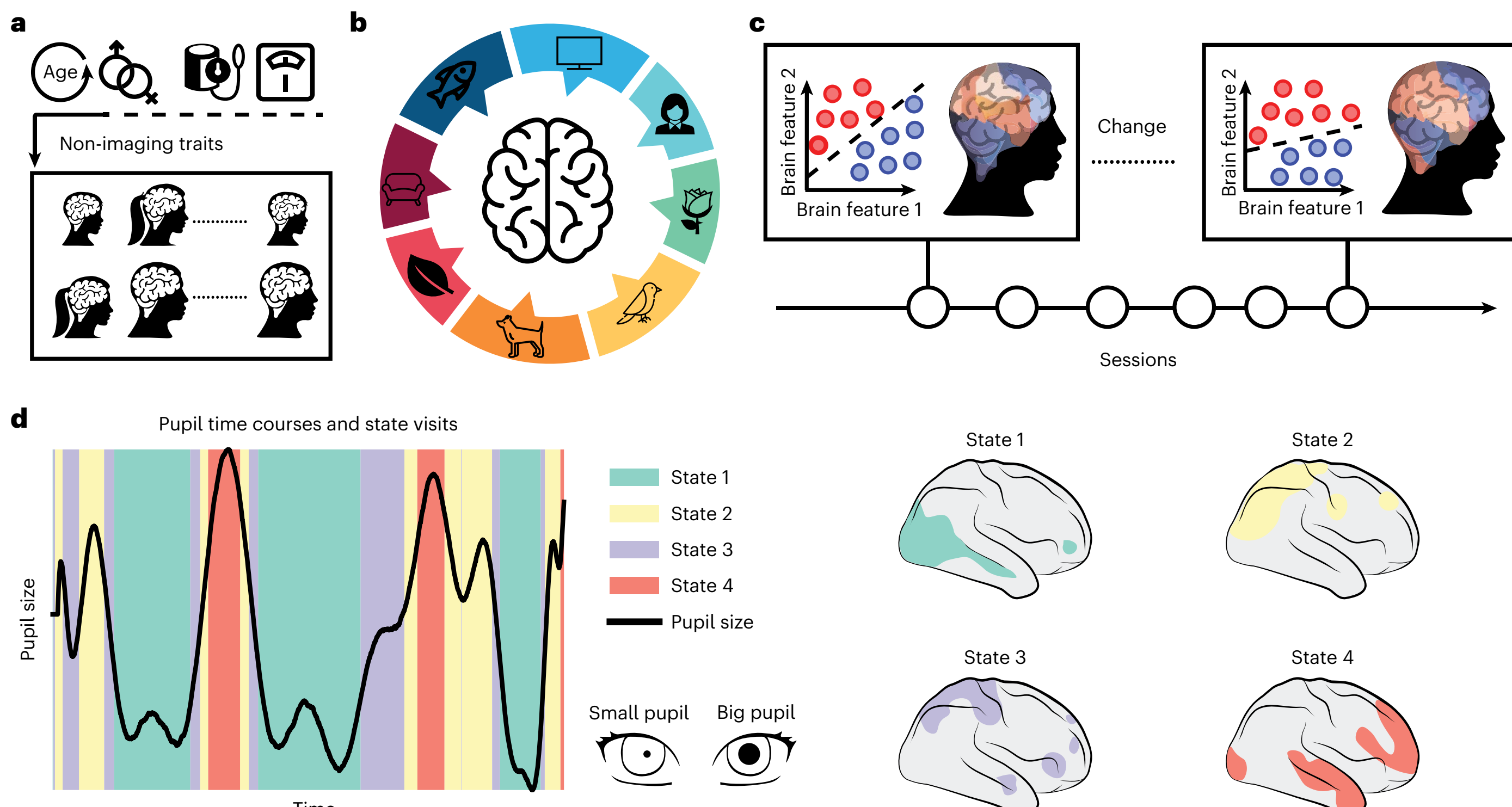

**Fig. 1 | Illustration of the four statistical tests. a**, The across-subjects test compares behavioral measurements (or traits) across multiple individuals to test subject trait differences. **b**, The across-trials test assesses differences in brain responses across experimental conditions, such as two types of stimuli. **c**, The across-sessions-within-subject test assesses changes in brain responses over experimental sessions, given an experimental paradigm such as the one used in **b**. **d**, The across-state-visits test assesses relationships between state time courses and concurrently recorded signals, where each state may correspond to the activation of a specific brain network.

where $s_t$ is a variable indicating which state is active at time point $t$, $\mu_k$ is the baseline activity for state $k$, $\beta_k$ represents the regression coefficients linking $X$ to $Y$ for state $k$ and $\Sigma_k$ is the covariance matrix for state $k$. This allows for flexible modeling of the data, where the parameters $\mu_k$, $\beta_k$ and $\Sigma_k$ may vary across states, remain global or not be modeled (in the case of the covariance matrix, this corresponds to using the identity matrix). This model reduces to the standard Gaussian HMM when $\beta_k$ is unmodeled and $\mu_k$ is state specific. Furthermore, the transition probabilities describe the likelihood of switching from one state to another:

$$P(s_t = k|s_{t-1} = l)$$

To estimate the posterior distribution of the model parameters, including the state time courses, represented as the probabilities $\gamma_{tk} = P(s_t = k|s_{t-1} = l, X_t, Y_t)$, the GLHMM uses variational inference.

Overall, thanks to its flexible parametrization, the GLHMM allows for many time-varying functional connectivity analyses, at the whole-brain level or targeting specific connections or networks[20].

### Performing statistical testing on the estimated model of brain dynamics

Once the model has been fitted to the data, we use formal statistical testing to examine the associations between the model parameters (representing different aspects of the time series' dynamics) and the behavioral or experimental variables. To assess whether these associations are statistically meaningful, the framework primarily relies on permutation-based inference, which does not impose any distributional assumption. This avoids issues when these assumptions are violated, which can lead to unreliable $P$ values and inflated false-positive rates. Alongside permutation-based methods, the framework also includes a test that uses structured Monte Carlo resampling, the across-state-visits test, which is discussed below.

As represented in Fig. 1, the four types of tests presented in this protocol are across-subjects, across-trials-within-session, across-sessions-within-subject and across-state-visits. We next succinctly describe the four tests, and further details can be found in the Supplementary Information.

The across-subjects test considers data from multiple individuals (or brain scans) to assess the associations between subject-specific model parameters encoding different aspects of brain dynamics on the one hand, and one or more subject-specific non-imaging traits (e.g., age, sex and/or cognitive capacity) on the other hand. For instance, we might be interested in testing the relationship between the time spent in the default mode network at wakeful rest and a clinical trait such as anxiety levels, cognitive decline or depression risk. For permutation testing, an important requirement is the exchangeability of subjects or scans, meaning that, after permutation, in the absence of a real effect, the distribution of the data remains the same as that of the unpermuted. However, if subjects have familial relationships, this assumption would be violated, making the test invalid. We address this issue in two ways. In the simplest scenario, the subjects or scans can be assigned to blocks, such that permutations are carried out at the block level, either within or between blocks. For example, if we had several scans per subject and one non-imaging trait per scan, the blocks would correspond to the subjects. In more complex scenarios, nested relationships between subjects or different types of relationships can be considered. Here, the user provides a hierarchical tree to account for this structure in the permutation scheme; for more details, see ref. 21.

The across-trials test considers experimental studies in which subjects perform a task across multiple trials within a single session or multiple sessions to assess differences in the states' time courses between experimental conditions (or subject actions). A typical example is a visual paradigm comparing two types of stimuli, where the goal is to identify when network activity significantly differs between conditions. The test generates a surrogate (null) distribution by performing permutations only on trials within the same session in which they were recorded. This test can be run at each time point throughout the trial to examine how the effects unfold over time.

The across-sessions-within-subjects test provides a new way to assess whether the brain–behavior relationship under study changes over slower time scales (i.e., over the course of multiple sessions). This approach can be used in longitudinal studies in which a subject is scanned repeatedly while performing a task involving one or more contrasts (e.g., stimuli or subject decisions). Unlike traditional approaches that shuffle trial data, this method operates at the level of regression coefficients. For each session, a regression model is fitted to estimate session-specific beta coefficients, which capture the relationship between brain activity (here, state time courses) and the experimental condition. To test for significant changes across sessions, the method generates a (null) distribution by randomly permuting these beta coefficients across sessions (rather than permuting the data). This approach accounts for differences in session length, variations in condition proportions and the lack of direct alignment between trials across sessions (as described in Supplementary Note 1). Similar to the across-trials test, this can be performed at each time point to produce a time-resolved statistical analysis.

The across-state-visits test, also novel in this context, evaluates whether the state time courses (represented by the Viterbi path, a discretized version of the state activation probabilities $\gamma_{tk}$) are associated with concurrently recorded physiological or behavioral signals over time, such as pupil size, heart rate or skin conductance. For instance, one state might correspond to an increase in pupil size, whereas another might correspond to a decrease. The Viterbi path thus serves as the contrast, enabling the comparison of differences in the second set of signals. Standard permutation methods are, however, not suitable for this test, because shuffling time points would disrupt the temporal structure of the data. Instead, using a Monte Carlo approach, we generate surrogate Viterbi paths that preserve the original transition timings but randomly reassign which states are visited at each transition in a structured manner. This is done in such a way that it maintains the statistical properties of the original data while breaking the observed association between states and the external signal; further elaboration on this test is available in Supplementary Note 2.

Together, these tests provide a framework for systematically studying dynamic brain-behavior relationships. The protocol below describes each step in detail, addressing the key challenges that users may encounter.

## Application of the method

The strength of the HMM framework comes from its ability to (1) detect fast changes in the properties of the data in a data-driven manner and (2) offer a cohesive representation of dynamics at both the group and the subject level by using a well-defined model. By leveraging this capacity, the HMM has been used to investigate several neuroscience questions in recent years, such as the nature of the sleep cycle, from fMRI data[10]; the long-term temporal structure of key brain networks during spontaneous cognition, by using MEG[8]; the relationship between the temporal patterns of whole-brain networks and the spontaneous replay of previously learned sequences in MEG[16]; the dynamics of memory retrieval throughout the cortex in fMRI[22]; the spectral characterization of large-scale cortical networks at rest in MEG[15]; and how fast-changing brain states relate to specific social behavior dynamics[23]. The comprehensive set of statistical tests introduced here can streamline and systematize the investigation of these and related questions, facilitating the exploration of associations between brain state dynamics and behavior.

However, the methods in this protocol are sufficiently general that they are not limited to neuroscience and can be applied to other fields. For instance, in economics, these may be used to identify periods when key political events correlate more strongly with economic indices, such as inflation, unemployment rates and gross domestic product. For example, during major political events like elections, shifts in government policies or international trade agreements, economic indicators may show increased correlation as businesses and consumers react to potential changes in regulation, taxation or trade relationships. Modeling these relationships can provide a better understanding of how such events influence consumer behavior and economic dynamics. In ecology, as another example, this protocol could be used to study animal migration patterns by identifying latent states that correspond to different stages of migration, such as foraging, resting or traveling. These states can be further analyzed to understand how they change in response to environmental factors such as food availability. For instance, shifts in migration routes or timing may be linked to climate change or human activities.

From a practical point of view, this protocol supports a range of industry-standard data formats, including CSV, text files and NIFTI, because of Python's robust data handling capabilities. Given this versatility, integrating data from other fields into the framework is straightforward and efficient. Documentation is available at 'Read the Docs' (https://glhmm.readthedocs.io/), including tutorials and examples.

In summary, the presented protocol has the potential to aid research in multiple fields involving temporal data by identifying latent factors underlying the dynamics of complex systems. The tests can also be applied generally to any time series, regardless of whether the HMM is used.

## Experimental design

As mentioned above, the statistics toolbox of this protocol includes four tests: across-subjects, across-trials, across-sessions-within-subject and across-state-visits. Figure 2 presents an overview of the procedure for applying these tests by using the toolbox. The procedure is divided into three parts: preparing the data, applying the statistical analysis and visualizing the results. Although part 1 includes Steps 2–5, which are specific to the HMM model, the statistical tests themselves can be applied to any data type and do not require HMM outputs.

### Install and set up the Python environment

Before beginning the analysis, we set up a Python environment and install the required packages. We start by creating a Python environment to manage dependencies separately from other projects. Once the environment is ready, we install the package by running the following command in the terminal:

```
pip install glhmm
```

This command clones the GitHub repository with all the code required for the procedure. The protocol can be run on a local computer or, if needed, on Google Colab.

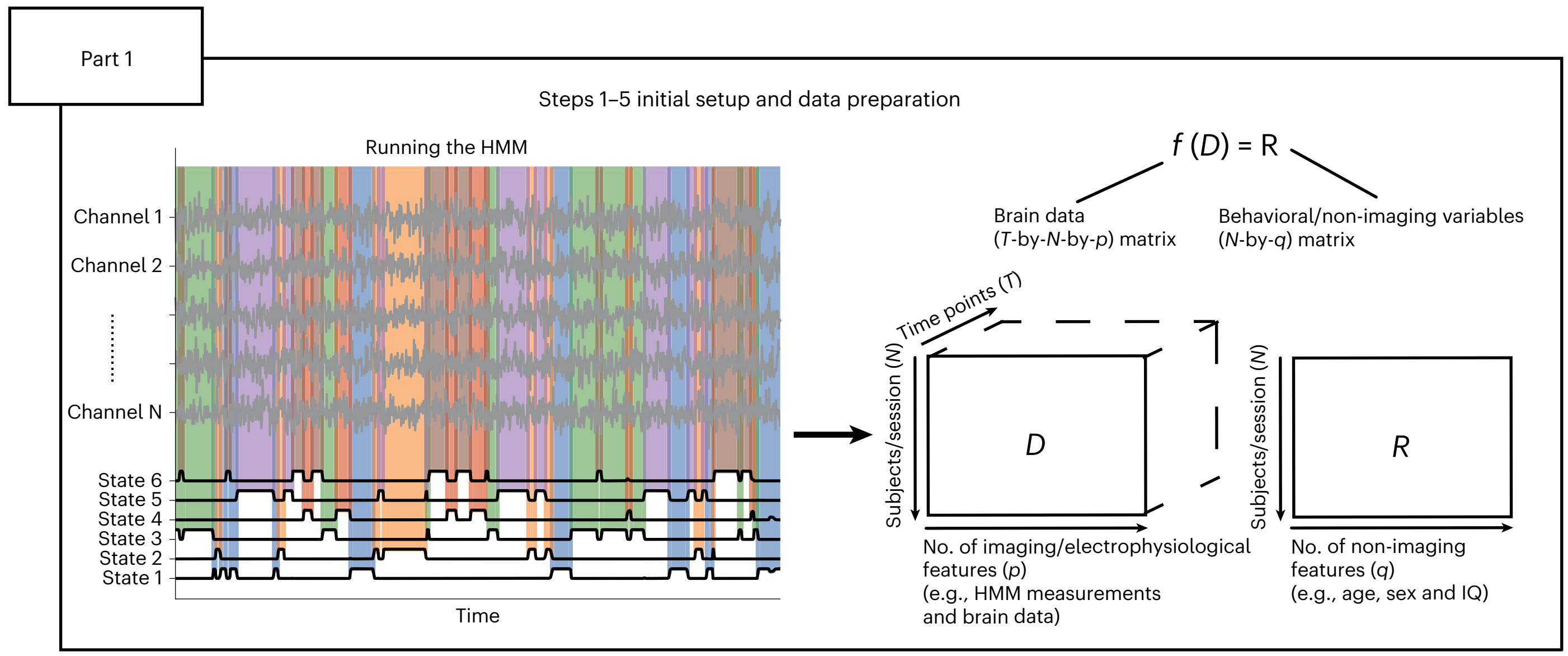

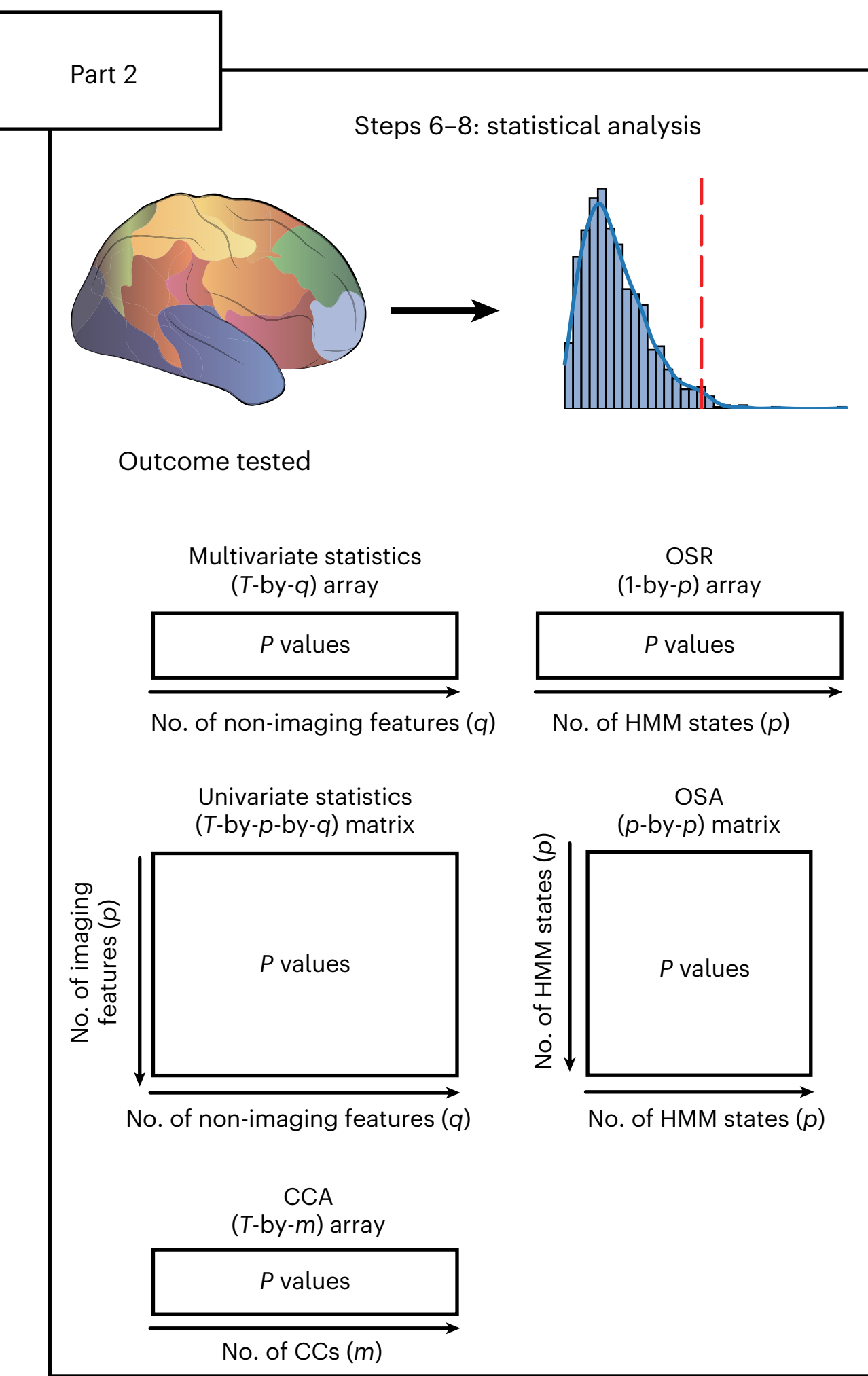

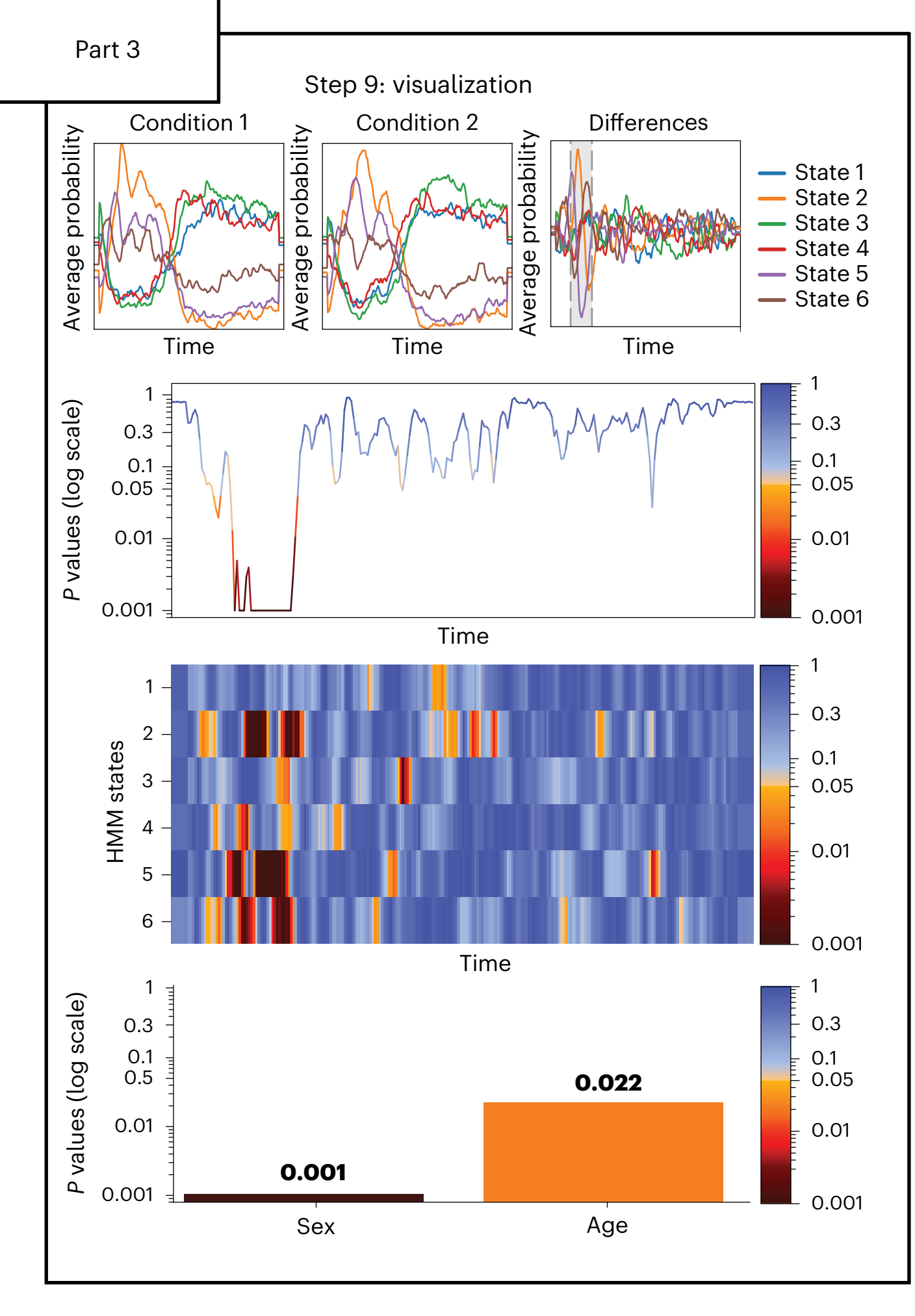

**Fig. 2 | Schematic of the analysis pipeline.** Part 1 (Steps 1–5), Brain data are optionally modelled using an HMM to estimate state time courses, which can be used to construct matrix *D* for statistical testing. Matrix *D* can also represent any data with the appropriate structure as defined in the protocol. Matrix *R* contains behavioral or other non-imaging variables. Part 2 (Steps 6–8), Statistical analysis between *D* and *R*, producing *P* value arrays whose structure depends on the chosen test. Part 3 (Step 9), Visualization of the results using different types of plots. CC, canonical correlation.

### Load data into the Python environment

To begin the analysis, we load brain and behavioral data. The aim is to examine potential associations between brain activity and behavior by using these two types of data. Functional brain data can come from different techniques, such as fMRI, EEG, MEG, LFP or ECoG (although the statistical tests are general enough to be directly applied to structural measures as well). Behavioral data or, more generally, non-imaging data can be cognitive or demographic information or any clinical variable. Although the brain data can go through extra processing by using the HMM model (described in Steps 2–5), these steps are optional. That is, any imaging set of variables, even if not produced by the HMM, can be used as *D* (Fig. 2).

### Data structuring for the HMM (optional)

When the brain data are prepared for training an HMM model, they need to be shaped as a [(No. of timepoints · No. of subjects/sessions) × No. of features] matrix. In this format, the data from all subjects or sessions are combined along the first dimension, while the second dimension represents the features, such as brain regions or channels. If the brain data are provided as a tensor (e.g., [No. of timepoints, No. of subjects, No. of features]), we can reshape them by concatenating time points and subjects or sessions along one dimension, with the features remaining as the second dimension.

### Preprocessing data (optional)

Before analysis, the raw data may need to be cleaned to remove noise and artifacts. The package offers tools for basic preprocessing, such as standardizing the data (to keep measurements on the same scale), filtering (for noise removal or to isolate specific frequency bands) and dimensionality reduction (using principal component analysis (PCA) or independent component analysis). If additional preprocessing steps are needed, these should be handled separately.

### Set up and train an HMM (optional)

The next steps are initializing and training the HMM with preprocessed data. Before training, the number of these states needs to be defined on the basis of the needs of the analysis and the size of the data[12,15,19]. In addition, the type of state model has to be chosen[7]. Once trained, the model saves the learned parameters and state time courses (referred to as 'gamma' in the code), which represent the probability of each state to occur at each time point. These state time course values are used in subsequent statistical tests to examine how state transitions relate to cognitive and behavioral measures.

Once the model is trained, the estimated parameters can be inspected to understand what each state represents and how the model behaves over time. These include the initial state probabilities (which reflect the state probabilities at the start of every segment of data), the average activation patterns (state-specific means), the covariance matrices (state-specific or state-averaged functional connectivity) and the transition probabilities (of transitioning from one state to another). The toolbox includes visualization tools for these elements, making it easier to evaluate the fitted model before proceeding to statistical analysis.

### Configure HMM outputs for statistical analysis (optional)

The HMM output takes different forms, depending on the type of test one wishes to carry out. By default, it produces continuous state time courses, which can be used to study changes over the full recording. Alternatively, the state time courses can be epoched, creating a 3D tensor, or summarized into a 2D matrix of aggregated statistics. When epoching is applied, the state time courses are divided into segments on the basis of specific experimental events (i.e., trials), such as responses to stimuli or other time-locked occurrences. This allows for the analysis of how brain states differ during these targeted periods. Finally, we can compute some form of aggregated statistics to generate a single set of values for each subject or session, summarizing the main patterns in the state time courses across the whole series or within specific time windows. These statistics include fractional occupancy (FO), which represents the proportion of time spent in each state during a given period; dwell time, the average duration spent in a state, reflecting its stability; switching rate, the frequency of transitions between states; and FO entropy, a measure of variability in the state visits, where high entropy indicates balanced state visits, and zero indicates that only one state is visited. Another possibility is to test specific parameters of the HMM, such as transition probabilities or specific state parameters.

**Table 1 | Options for statistical testing**

| Characteristic | Types of tests |
|---|---|
| Input data | Brain data (*D*)[a]<br>Behavioural data (*R*)[a]<br>Viterbi path (*D*)[b]<br>Non-imaging signals (*R*)[b] |
| Permutation/parametric testing | Parametric testing if the number of permutations is set to 0 |
| Methods supported | Multivariate<br>Univariate<br>CCA<br>One-state-versus-the-rest[b]<br>One-state-versus-another-state[b] |
| Category identification | Automatically detects data type in (*R*) and applies:<br>Independent *t* test (Boolean), ANOVA (categorical), *F*-regression (continuous, multivariate), Pearson correlation *t* test (continuous, univariate); default: false |
| Test combination | Supports NPC on *P* values across rows, columns or both; default: false |
| Confounding variables | Regresses out confounding effects from *D* and *R*; default: none<br>Regresses out confounding effects from *R*[b] |
| Handling subject dependencies | Hierarchical permutations for family relationships[c] |
| Multiple testing correction and cluster statistics | Supports classical multiple-comparison corrections (e.g., Bonferroni, Benjamini–Hochberg), FWER (e.g., MaxT) and cluster-based statistics (spatial/temporal) |
| Output | Dictionary with *P* values, base statistics, test types, methods used and correction details |

[a]Across-subjects, across-trials and across-sessions only. [b]Across-state-visits only. [c]Not applicable for across-trials, across-sessions or across-state-visits.

For the purposes of this protocol, any form of data *D*, whether it is continuous, epoched or aggregated, is considered to originate from the HMM and is referred to as 'brain data'. However, *D* does not need to be a product of the HMM; it can represent any measure as long as it has the correct structure. When *D* is structured as a 3D tensor with dimensions [No. of timepoints × No. of subjects or sessions × No. of states or features], statistical tests can be performed for each time point to analyze the temporal aspects of the data. Alternatively, if *D* is structured as a 2D matrix with dimensions [No. of subjects or sessions × No. of channels or features], statistical testing is performed on temporally aggregated data. The behavioral matrix *R*, which will be tested against *D* as shown in Part 1 of Fig. 2, has dimensions [No. of subjects or sessions × No. of behavioral features]; these features can include any non-imaging variables such as cognitive capacity, age, sex or the experimental condition.

In addition to generating summary statistics for analysis, the fitted HMM can also be inspected directly. Users can access the estimated state means, covariance matrices and transition probabilities to examine the spatial and temporal properties of each state. These outputs support model interpretation and quality control before proceeding to statistical testing.

### Statistical analysis

All the settings needed to perform these statistical tests are listed in Table 1. These, unless obvious, will be explained next.

**Types of tests**

Permutation testing, the primary method used in the framework, is a non-parametric approach that shuffles the data to generate a surrogate distribution in which the key property that we want to test (and not others) is broken. However, the toolbox also allows for parametric testing, which is computationally much faster and applicable on small samples and when assumptions hold reasonably.

Another feature of the framework is its ability to handle missing values in the dataset. During the analysis, tests automatically exclude these missing values so that incomplete data do not interfere with calculations or affect result reliability. This approach assumes that missing values occur completely at random. If the missingness follows other patterns, such as missing at random or missing not at random, this approach may produce biased or even invalid results, depending on the extent and nature of the missing data.

As stated, this protocol supports the following types of tests: across-subject, across-trials, across-sessions-within-subject and across-state-visits.

**Methods supported**

The choice of method depends on the specific research question and data structure. The statistical measures used to assess the relationship between the matrices $D$ and $R$ derive from different functions (e.g., regression metrics and correlation coefficients). In all cases, the null hypothesis is that there is no association between the brain data $D$ and the behavioral data $R$. Within this framework, $D$ is structured as $N \times p$, and $R$ is structured as $N \times q$, where:

- $N$ = number of observations (e.g., subjects or trials),
- $p$ = number of predictors (e.g., features in $D$), and
- $q$ = number of outcomes (e.g., behavioral variables in $R$) being tested.

For across-subjects, across-sessions-within-subject and across-trials-within-session tests, the protocol provides multivariate regression tests and univariate tests as well as canonical correlation analysis (CCA) (See Part 2 in Fig. 2).

Multivariate regression tests examine the overall relationship between $D$ and each variable or outcome in $R$. This approach produces $P$ values, one for each outcome in $R$. For example, if $R$ represents 12 HMM states and $R$ includes two behavioral variables like sex and age, the output contains two $P$ values, one for each behavioral variable. The setup can also be reversed by treating $R$ as the independent variable and $D$ as the dependent variable, with the former case being the default. Multivariate tests use $F$ statistic as the default base statistic for permutation testing. To assess the predictors' contribution to the prediction, it also returns regression coefficients and individual $P$ values per regressor or predictor (similar to those derived from $t$ tests in multiple linear regression).

Univariate tests independently assess the relationship between each feature in $D$ and each variable in $R$. When using the same example, the output is a $12 \times 2$ matrix of $P$ values, where each element reflects the association between a specific predictor in $D$ and an outcome in $R$. The default base statistic for univariate tests is the $t$-statistic derived from Pearson correlation.

CCA provides a single $P$ value summarizing the overall relationship between the variables in $D$ and $R$, capturing how brain states in $D$ relate to the behavioral measures in $R$. By default, the analysis includes one CCA component, but users can specify a different number of components if desired.

For across-state-visits tests, in which we assess the relationship between state time courses and another simultaneously collected set of time series, the protocol includes two additional methods: one-state-versus-the-rest (OSR) and one-state-versus-another-state (OSA). Here, $D$ is given as the Viterbi path (such that $p$ is the number of states), that is, the most likely sequence of states over time, with each time point categorically assigned to one state; and $R$ represents, for example, a set of physiological time series (such as pupil size or skin conductance). Assuming that $R$ has a single column for simplicity, in OSR tests the mean value of $R$ for a specific state is compared to the mean value of $R$ across all other states. By default, the test evaluates whether the mean $R$ for the specific state is larger than the average of the remaining states. This produces $p$ $P$ values. In OSA, the mean values of $R$ are compared between all possible pairs of states, generating a $p$-by-$p$ matrix of $P$ values. Each comparison is based on the difference in the mean value of $R$ between two states.

### Statistical analysis: test combination (optional)

For the across-subjects, across-sessions-within-subject and across-trials-within-session tests, the protocol includes the non-parametric combination (NPC) algorithm to combine multiple *P* values into fewer *P* values with increased statistical power[24,25]. Specifically, instead of getting a *P* value for each pair of variables—that is a $(p \times q)$ matrix of *P* values—the NPC algorithm condenses these into one *P* value per row ($1 \times p$ *P* values), one *P* value per column ($1 \times q$ *P* values) or a single *P* value for the entire test.

In our implementation, we use Fisher's method as the combining function, which efficiently aggregates the *P* values while maintaining sensitivity to small values. This approach differs from CCA, which also produces a single *P* value but does so by testing the strength of a multivariate relationship between two variable sets. In addition to statistical inference, CCA provides a latent representation of the data in the form of canonical variables—linear combinations that maximize correlation between the sets. NPC, by contrast, aggregates test results across multiple comparisons. Although it offers flexibility and interpretability, it does not yield a latent representation or model the joint multivariate structure directly.

### Multiple testing correction and cluster statistics

When performing statistical tests, we need to correct for multiple testing correction to control false positives or type 1 errors. The protocol includes standard correction methods from the statsmodels module, such as Bonferroni and false discovery rate (FDR) control using the Benjamini–Hochberg procedure. In addition, it supports family-wise error rate (FWER) correction with the MaxT method[26].

For data with spatial or temporal structure, the protocol also includes support for cluster-level inference[27]. A cluster is a contiguous group of tests that survive a predefined statistical threshold. Clusters can be formed by multiple neighboring voxels or consecutive time points, depending on the type of analysis. The test statistic for a cluster can be its size (called cluster extent) or the sum of the test statistics within it (called cluster mass). The significance of each cluster is assessed by comparing its test statistic to the distribution of the maximum test statistic across all clusters. This distribution is obtained through permutation testing, and because it is based on the maximum statistic, the *P* values are FWER-corrected for multiple testing at the cluster level.

### Visualizing statistical results

Finally, the protocol includes steps for visualizing and interpreting results in a way that is both clear and easy to understand, by using various graphical tools like heatmaps, bar graphs and line plots to display the *P* values; Part 3 in Fig. 2 shows some examples. To highlight significant differences, we use a color map in a logarithmic scale that shifts from dark red to yellow where there is significance, and from gray to blue where there is not.

## Comparison with other methods

The present protocol is designed to accommodate both task-based experimental designs and resting-state experiments. Although many existing toolboxes such as FSL[28], SPM[29], AFNI[30], MNE[31] and CONN[32] provide robust support for standard group-level analyses, they are often limited to time-averaged representations of brain activity. Our protocol addresses this limitation by providing statistical inference on time-varying features (e.g., brain dynamics) and their relationship to behavioral or physiological variables.

What we refer to as across-subjects and across-trials analyses are supported by all the other toolboxes, enabling group-level inference and condition-based contrasts. However, support for longitudinal analyses in the way we present here (through the so-called across-sessions-within-subject testing) is lacking. Some tools (e.g., AFNI and CONN) allow users to combine multiple sessions by summarizing each session separately (e.g., by computing average connectivity per session) and then comparing those summary metrics by using group-level statistics. By contrast, our protocol retains the full temporal structure across sessions and enables trial-by-trial or time point-by-time point inference. This makes it possible to analyze how brain dynamics change from

**Table 2 | Comparison of statistical testing features across toolboxes**

| Category | Feature | FSL | SPM | MNE | AFNI | CONN |
|---|---|---|---|---|---|---|
| Type of test | Across-subjects | Yes | Yes | Yes | Yes | Yes |
| | Across-trials | Yes | Yes | Yes | Yes | Yes |
| | Across-sessions within-subject | No | No | No | No | No |
| | Across-state-visits | No | No | No | No | No |
| Methods supported | Multivariate | Yes | Yes | Yes | Yes | Yes |
| | Univariate | Yes | Yes | Yes | Yes | Yes |
| | CCA | No (default); yes via PALM | No (default); yes via PALM | No (default); yes via PALM | No | No |
| | One-state-versus-the-rest | No | No | No | No | No |
| | One-state-versus-another-state | No | No | No | No | No |
| Test combination | Combine tests across rows/columns/full matrix | No (default); yes via PALM | No (default); yes via PALM | No (default); yes via PALM | No | No |
| Hierarchical permutation | Account for family structure | No (default); yes via PALM | No (default); yes via PALM | No (default); yes via PALM | No | No |
| Multiple testing correction and cluster statistics | FWER | Yes | Yes | Yes | Yes | Yes |
| | FDR | Yes | Yes | Yes | Yes | Yes |
| | Cluster-based statistics | Yes | Yes | Yes | Yes | Yes |

Each row represents a specific feature or analysis type, with 'Yes' indicating that the toolbox supports the functionality and 'No' indicating that it does not support the functionality.

session to session and to ascertain when these changes take place as opposed to just whether they do. In addition, our protocol includes across-state-visits testing, which allows users to link moment-by-moment occurrences of brain states to concurrently recorded behavioral or physiological variables (e.g., pupil size or heart rate), a type of analysis not supported by the other toolboxes.

All of the compared tools support univariate and multivariate testing. However, CCA, a method well suited for linking multivariate neural features with multivariate behavioral data, is not natively supported in FSL, SPM, MNE, AFNI or CONN. In some cases (e.g., FSL, SPM and MNE), CCA can be added by using external tools such as Permutation Analysis of Linear Models (PALM). By contrast, CCA is directly implemented in our framework.

The toolbox also supports combining results from multiple related tests, such as different cognitive measures, into a single test. This makes it possible to test whether patterns in the brain are linked to a broader behavioral profile rather than looking at each variable in isolation. Although PALM supports test combination and could be used alongside other packages, this typically requires a manual setup. By contrast, our protocol integrates test combination directly into its core workflow, making it more accessible and easier to apply. Taking inspiration from PALM, our framework also provides hierarchical permutation testing to account for family relationships between subjects.

All toolboxes provide standard procedures for multiple testing correction (e.g., FWER and FDR) and cluster-based statistics. In our framework, these procedures are built directly into the main analysis pipeline, making them easier to apply without additional configuration.

Table 2 summarizes the core differences in statistical testing capabilities across these toolboxes.

## Expertise needed to implement the protocol

The toolbox described in this protocol is designed to be easy to use for practitioners with varying levels of programming experience, although some basic familiarity with Python is required. Although extensive expertise in statistical methods is not required, users should have some ability to interpret the results appropriately. To facilitate its application, the protocol includes clear documentation, example datasets and tutorials for each of the four statistical test designs. Each tutorial has step-by-step instructions with practical examples, so users do not have to write code from scratch. This allows users to learn how to train an HMM model, select the appropriate data for input, interpret the results and draw meaningful conclusions with minimal time investment.

### Limitations

The presented toolbox exclusively uses linear models for statistical testing. This can be seen as a limitation when the relationships in the data are nonlinear. However, the linear methods presented can easily be extended to the nonlinear case by using an appropriate basis expansion[33].

## Materials

### Data

This protocol outlines pipelines for a comprehensive set of statistical tests, applicable to a broad range of scientific questions in neuroscience. We demonstrate these tests by using publicly available data, as summarized below. In Protocol 1 (across-subjects), we analyze resting-state brain activity from 1,001 Human Connectome Project (HCP) participants across four sessions to examine its relationship to 15 traits related to cognitive performance (Supplementary Table 1). In Procedure 2 (across-trials), we study MEG data from a single person who participated in 15 sessions. During each session, the person watched both animate and inanimate objects while their brain activity in the occipital lobe was recorded. This analysis assesses differences in the brain responses when the person looks at animate objects compared to inanimate ones. In Procedure 3 (across-sessions-within-subject), we use the same dataset as in Procedure 2, but this time focus on changes over multiple sessions. This analysis shows whether the person exhibits changes in stimulus processing over time (i.e., across sessions) due, for example, to learning, or whether their brain representations remain stable. In Procedure 4 (across-state-visits), we analyze MEG data from 10 participants scanned at rest in a dark room. During the scans, pupil size and brain activity were measured concurrently. Nine participants completed two sessions, and one completed a single session. This analysis explores how changes in brain states, like the default mode network, relate to variations in pupil size[34,35]. All data (except HCP) needed to reproduce the results of these workflows are hosted on Zenodo (https://doi.org/10.5281/zenodo.15213970), and the code is available at GitHub.

### Software

- Computer requirements. Any personal computer, Mac or Linux computer can be used to run this protocol.
  ▲ **CRITICAL** If a local computer is unavailable, the protocol can run via Google Colab for free. For this, a computer with a stable internet connection and a modern web browser such as Chrome or Firefox are required.
- Python installation. Download and install Python from the official website: https://www.python.org/downloads/.
  ▲ **CRITICAL** Make sure to install the version compatible with the GLHMM package requirements. See https://github.com/vidaurre/glhmm.
- Recommended tools. To manage Python packages and environments effectively, we recommend using Anaconda, Spyder or Visual Studio Code. Anaconda simplifies package management and environment setup, while Visual Studio Code provides a robust development environment with useful extensions.
- GLHMM Python package. Install the GLHMM Python package and its dependencies by using pip. The package is available for download at https://github.com/vidaurre/glhmm. This protocol is based on the latest release of GLHMM (version 1.1.1, released in July 2025). The GLHMM toolbox is available both as a Python package and as a GUI. The GUI allows users to run analyses through a user-friendly interface. A video tutorial (about 30 min) demonstrating how to set up and use the GUI is also available; the link is provided in the GitHub repository.

### Input data

- Temporal brain data. These continuous data can correspond to any neuroimaging modalities, such as fMRI, MEG, EEG, ECoG or LFP.

▲ **CRITICAL** Other types of temporal (or, more generally, sequential) datasets can also be used as previously mentioned. Except for the initial steps related to the fitting of the HMM, the across-subject tests can also be applied to structural (i.e., non-functional) data.

- Behavioral measures. These are additional inputs used in statistical analysis alongside the temporal data. These behavioral measures, or more generally, non-imaging traits, can include both discrete and continuous variables, such as age, sex, gender, race, ethnicity, IQ, demographic information and experimental events like responses to stimuli relevant to the study.

## Experimental setup

### Install and set up the Python environment

● **TIMING** 1–5 min

To set up Python and manage the required tools and packages, we recommend creating a conda environment. If Anaconda is not installed, it can be downloaded from the Anaconda website. After installing Anaconda, create and activate a new environment by running the following commands:

```
conda create -- name glhmm_env python =3.10
conda activate glhmm_env
```

Once the environment is activated, install the GLHMM package via pip:

```
pip install glhmm
```

This command installs the package along with all required dependencies. No additional setup is needed.

The Jupyter notebooks used in this protocol are available on the GLHMM Protocols GitHub repository. These notebooks contain detailed examples for each of the four procedures described in this paper. Required datasets can be downloaded from Zenodo. However, the notebooks are designed to automatically download the data if they are not already present in the expected directories.

▲ **CRITICAL** After setting up the Python environment, load the necessary libraries required to run the protocols:

```
import numpy as np
import matplotlib.pyplot as plt
import pickle
from pathlib import Path
from glhmm import glhmm, graphics, statistics, io, preproc
```

These libraries provide tools for data loading, preprocessing, statistical analysis and visualization.

## Procedure 1: across-subject testing

● **TIMING** 0–2 min

1. Load and prepare data. Load data into the Python environment. For Procedure 1, we use data from the HCP Young Adult study[36]. Specifically, we work with resting-state fMRI data from 1,001 participants, each with 50 parcellations/channels identified through independent component analysis. Each participant's data contain 4,800 time points across four sessions (1,200 time points per session, each lasting about 15 min). These data are stored in `data_measurement_HCP.npy` and loaded into the variable `D_raw`, referred to as 'matrix *D*'. Behavioral data for 15 cognitive traits related to fluid intelligence are stored in `data_cognitive_traits_HCP.npy` and loaded into `R_data`, referred to as 'matrix *R*'.

See Supplementary Table 1 for the full list of traits. Additional confounds, such as sex, age and fMRI head motion, are included in `confounds_HCP.npy`. We load the data using Numpy:

```
D_raw = np.load("data_HCP / data_measurement_HCP.npy")
R_data = np.load("data_HCP / data_cognitive_traits_HCP.npy")
confounds = np.load("data_HCP / confounds_HCP.npy")
```

In summary:

- `D_raw`: [4800, 1001, 50]—4,800 time points, 1,001 subjects or sessions and 50 features.
- `R_data`: [1001, 15]—1,001 subjects and 15 features (e.g., cognitive traits).
- `confounds`: [1001, 3]—1,001 subjects and 3 confounding variables.
  ▲ **CRITICAL** If using your own data, make sure that they are structured similarly and stored as Numpy arrays.

● **TIMING** <10 s

2. Load and prepare data: data structuring for the HMM. When training the HMM, we need to format the data as a 2D matrix of shape [(No. of timepoints · No. of subjects) × No. of features]. This means combining the time points from all subjects into one continuous sequence while keeping features (e.g., brain parcellations in this case) in the second dimension. For example, our dataset *D* is shaped like [4800, 1001, 50] ([No. of timepoints × No. of subjects × No. of features]), it needs to be reshaped to [4804800, 50]. We use the `get_concatenate_subjects` function to reshape the data by concatenating the time pointsand the `get_indices_timestamp` function to create indices marking where each subject's data start and end:

```
D_con = statistics.get_concatenate_subjects(D_raw)
idx_subject = statistics.get_indices_timestamp(
  D_raw.shape [0],
  D_raw.shape [1])
)
```

The generated indices for each session will look like this:

```
[[0 4800]
[4800 9600]
...
[4795200 4800000]
[4800000 4804800]]
```

● **TIMING** <10 s

3. Load and prepare data: preprocessing data. Because the data from the HCP are already cleaned and preprocessed when downloaded, the next step is to standardize the full time series before performing further analysis. Standardizing is important, especially when comparing data between different individuals, because it helps to ensure that the analysis is not affected by noise or differences in measurement scales. To standardize the data, we use the `preprocess_data` function from the module `preproc`. This function has many options for processing data, but here we focus on standardization. Standardization makes sure that each signal has an average (mean) of 0 and an s.d. of 1. Here is how to do it:

```
D_preproc, idx_preproc = preproc.preprocess_data(
  data = D_con,
  indices = idx_subject,
  standardise = True
  )
```

By entering the concatenated data (`D_con`) and the indices of each subject (`idx_subject`) and setting `standardise = True`, the function standardizes the whole dataset.

● **TIMING** <4 h

4. Load and prepare data: set up and train an HMM. To start using GLHMM, the first step is to set up the GLHMM model and choose the right settings. For a standard Gaussian HMM, we are not focused on interactions between different sets of data, so we set `model_beta ='no'`. In this example, the number of states is controlled by the parameter *K* and is set to 12. Each state is represented as a Gaussian distribution with its own unique average (mean) and full covariance structure, meaning that each state has a distinct pattern. To set this up, we set `covtype= 'full'`, and the model handles the state-specific mean automatically. Here is how to initialize the GLHMM model:

```
K = 12
hmm_HCP = glhmm.glhmm(model_beta ='no', K=K, covtype ='full')
```

Once the model is initialized, it is time to train the HMM by using the preprocessed data `D_preproc` and the subject index matrix idx_subject. In this case, we are not modeling interactions between two different time series, so we set X=None. The Y input should be the preprocessed time series data (`D_preproc`) that we want to use for estimating states.

```
Gamma, Xi, FE = hmm_HCP.train(X=None, Y=D_preproc, indices= idx_subject)
```

The trained model returns `Gamma` (the state probabilities at each time point), `Xi` (the joint probabilities of past and future states conditioned on the data) and `FE` (the free energy of each iteration).
After training, the learned state properties and transition dynamics, such as initial state probabilities, state means, covariances and transition probabilities, can be inspected to evaluate the model. These visualizations are included in the accompanying notebook.

● **TIMING** <40 s

5. Load and prepare data: configure HMM outputs for statistical analysis. To prepare for statistical analysis, we calculate the aggregated summary statistics from the Gamma values. For each subject, we compute FO, which represents the probability distribution of time spent in each state. FO shows how much time a subject spends in each of the 12 states across their recording. These values are our brain data input for the statistical test for each subject and are stored in the variable `D_fo`, which is generated by using the following code:

```
D_fo = glhmm.utils.get_FO(Gamma, idx_subject)
```

The resulting matrix has dimensions [1001, 12], where each row corresponds to a subject, and each column represents a state. The values in each row sum to 1 and provide a normalized summary of the time spent in each state.

● **TIMING** <1 s

6. Statistical analysis: types of tests. For statistical testing, we use the `test_across_subjects` function from the `statistics.py` module to test the relationship between `D_fo` (brain data, *D*) and `R_data` (behavioral measurements, *R*) for each subject.
▲ **CRITICAL** The `test_across_subjects` function assumes that all subjects can be permuted without affecting the results, which is known as being exchangeable. However, in practice, some subjects may be related, which violates this assumption. To handle this, we need to use an Exchangeability Block (EB) file to organize subjects into family blocks such that any permutations of the data respect family structures.

## BOX 1

## Details of the results dictionary

The result test dictionary stores the outputs of statistical tests and contains:
- 'pval': *P* values computed under the null hypothesis
- 'base_statistics': the observed test statistic calculated from the original (unpermuted) data.
- 'null_stat_distribution': test statistics generated under the null hypothesis, where the first row corresponds to the observed test statistic ('base_statistics').
- 'statistical_measures': dictionary specifying the type of test statistic in each column in ('base statistics'), such as *t*-statistics or *F*-statistics.
- 'test_type': type of test performed (e.g., across-subject test).
- 'method': analytical approach used (e.g., multivariate and univariate).
- 'combine_tests': indicates whether the NPC method was applied to summarize *P* values.
- 'max_correction': whether Max-statistic correction was used for multiple comparisons.
- 'Nnull_samples': total number of null samples including the observed one.
- 'test_summary': dictionary summarizing the test results.
- 'pval_f_multivariate': *F* test *P* values for multivariate tests and Nnull_samples > 1.
- 'pval_t_multivariate': *t* test *P* values for multivariate tests and Nnull_samples > 1.

- *Creating the EB.csv file*. The EB.csv file organizes subjects into family blocks to maintain these structures during permutation testing. You can specify the file location as shown below:

```
# Exchangeability Block (EB) information
dict_fam = {
'file_location': 'EB.csv'
}
```

A step-by-step tutorial for creating an EB.csv file for the HCP dataset is available in ref. 21.

● **TIMING** <10 min

7. Statistical analysis: methods supported. With the EB.csv ready, the next step is to set up and run the `test_across_subjects` function. Below is an example of how to configure the required inputs and perform a multivariate statistical test with 10,000 permutations. For this analysis, we use brain data (`D_fo`) and behavioral measurements (`R_data`) as inputs. Additional parameters include confounds (`confounds_data`), the number of permutations (`Nnull_samples`), the analysis method (`method`) and the family dictionary (`dict_family`).

```
# Set parameters for multivariate testing
method = "multivariate"
Nnull_samples = 10_000 # Number of permutations
# Perform multivariate analysis
result_multivariate = statistics.test_across_subjects(
  D_data =D_fo,
  R_data = R_data,
  confounds= confounds_data,
  Nnull_samples= Nnull_samples,
  method = method,
  dict_family = dict_fam,
  )
```

The results of the test are stored in the `result_multivariate` dictionary, which contains detailed results, including *P* values, test statistics and baseline measures. For a breakdown of the dictionary structure, see Box 1.

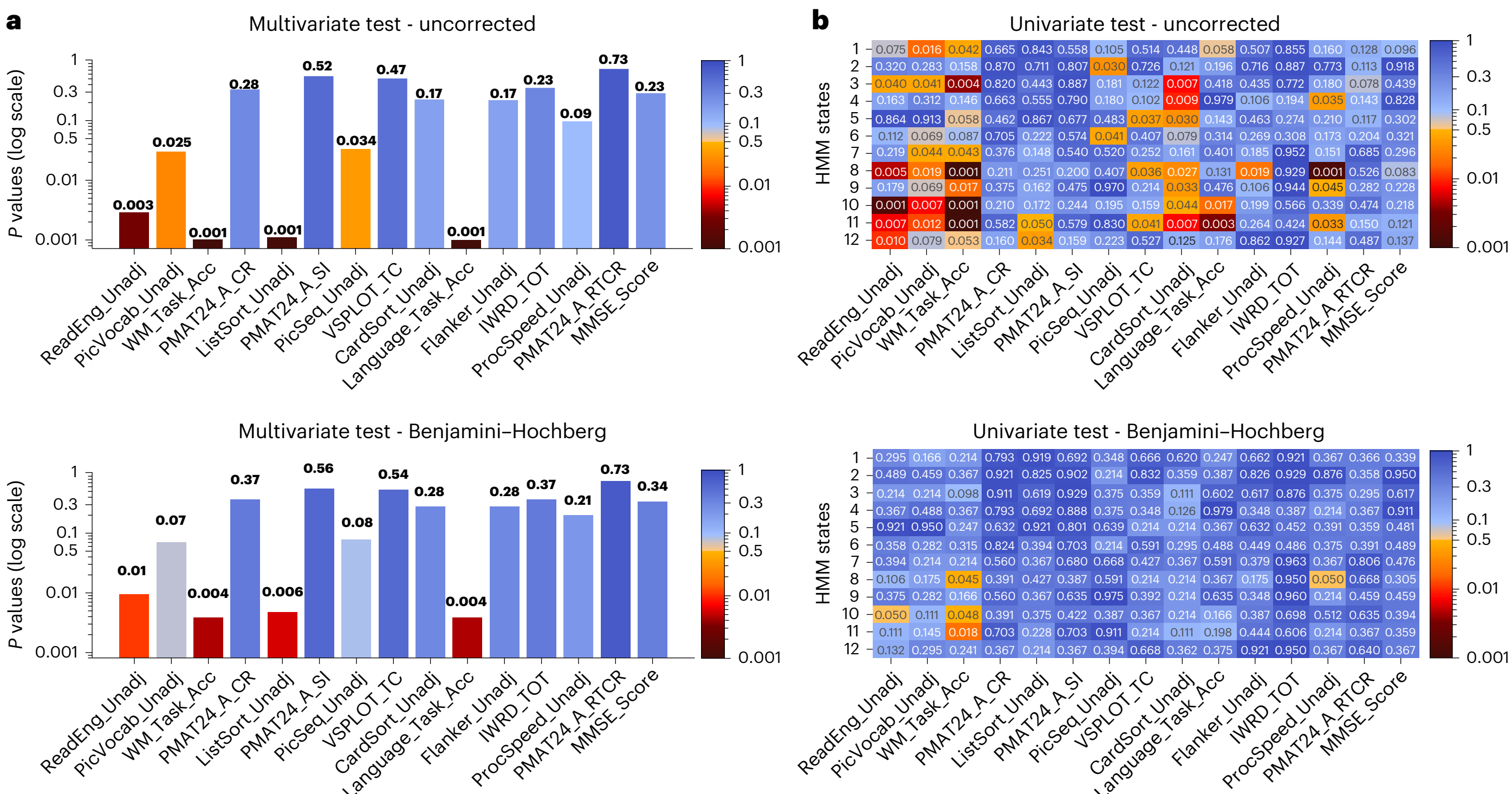

**Fig. 3 | Result from Procedure 1. a**, Results of the multivariate tests without and with the Benjamini–Hochberg corrected *P* values. **b**, Results of the univariate tests with the same settings.

● **TIMING** <5 s

8. Statistical analysis: multiple testing correction and cluster statistics. To minimize the risk of false positives in statistical tests, we can apply FWER correction by using the MaxT method. This is done during permutation testing by setting `FWER_correction =True` in the `test_across_subjects` function. This adjustment accounts for multiple testing correction by modifying the significance levels. If a different correction method, such as Bonferroni or Benjamini–Hochberg, is preferred, we can use the `pval_correction` function to adjust the *P* values after running the test. In this example, we apply the Benjamini–Hochberg method by providing the *P* values from `result_multivariate` and setting `method='fdr_bh'`:

```
pval_corrected, rejected_corrected = statistics.pval_correction(
result_multivariate,
method ='fdr_bh'
)
```

The function returns two outputs: `pval_corrected`, which contains the adjusted *P* values, and `rejected_corrected`, a Boolean array indicating which hypotheses are rejected on the basis of the corrected *P* values.

● **TIMING** <5 s

9. *Visualization: visualizing statistical results*. Visualizing both uncorrected and corrected *P* values helps to identify significant findings before and after applying corrections. Figure 3 displays results from multivariate and univariate tests. Although this demonstration focuses on multivariate testing, running a univariate test is straightforward. Set `method="univariate"` when running the test. The visualizations include bar plots for multivariate tests and heatmaps for univariate tests. These functions are part of the graphics module. For the different plots, we use `alpha = 0.05` to highlight *P* values below this threshold:

`# Features of cognitive traits` (see Supplementary Table 1 for details)

```
features = ['Read Eng_Unadj', 'PicVocab_Unadj', 'WM_Task_Acc', 'PMAT
24 _A_CR', 'ListSort_Unadj', 'PMAT 24 _A_SI', 'PicSeq_Unadj',
'VSPLOT_TC', 'Card Sort_Unadj', 'Language_Task_Acc','
  Flanker_Unadj', 'IWRD_TOT', 'ProcSpeed_Unadj', 'PMAT 24 _A_RTCR',
  'MMSE_Score'
]
alpha = 0.05 # Threshold for the p- value plots graphics.
plot_p_values_bar(
  result_multivariate["pval"],
  title_text=" Multivariate Test - Uncorrected",
  alpha = alpha,
  xticklabels= features,
  xlabel_rotation =45,
)
# Plot corrected p- values graphics. plot_p_values_bar(
  pval_corrected,
  title_text=" Multivariate Test - Benjamini - Hochberg",
  alpha = alpha,
  xticklabels= features,
  xlabel_rotation =45,
)
```

The bar plots display *P* values for each feature, with corrected values shown by using the Benjamini–Hochberg method. For univariate tests, use the `plot_p_value_matrix` function to generate a heatmap of *P* values.

## Procedure 2: across-trials testing

● **TIMING** <1 min

▲ **CRITICAL** For Procedure 2, we analyze MEG data collected from a single person who participated in 15 experimental sessions over about 6 months. During each session, the participant engaged in multiple trials and viewed animate and inanimate objects.

1. Load and prepare data: load data into the Python environment. The data include 72 MEG channels around the occipital lobe, stored in `D_raw.pkl`. This is a list in which each element corresponds to a session. These data are loaded into the variable `D_raw`, referred to as 'matrix *D*'. Behavioral data for the 15 sessions are stored in `R_data.pkl` and loaded into the variable `R_data`, referred to as 'matrix *R*'. This list contains trial-level information indicating whether the stimulus presented was an animate or inanimate object. This enables testing whether the brain responds differently to these conditions. To perform epoch-based analyses, we also load a list of event markers stored in `event_markers.pkl`. These markers include information about the time points at which stimuli were presented and are loaded into the variable `event_markers`. We use the pickle module to load the data:

```
with open("D_raw.pkl", "rb") as f: D_raw = pickle.load(f)
with open("R_data.pkl", "rb") as f: R_data = pickle.load(f)
with open("event_markers.pkl", "rb") as f: event_markers = pickle.
load(f)
```

In summary:
- D_raw: MEG data as a list of 2D matrices, where each matrix represents a session and has the shape [No. of time points × No. of channels] for that session.
- R_data: behavioral data as a list of arrays, where each array corresponds to a session and encodes trial information (0 for inanimate objects and 1 for animate objects).
- event_markers: a list of arrays, where each array corresponds to a session and includes the stimulus presentation time points and metadata.

● **TIMING** <10 s

2. Load and prepare data: data structuring for the HMM. For HMM training, we must format the MEG data to a 2D matrix with the shape [(No. of time points across all sessions) × No. of channels]. This involves concatenating the data from all 15 sessions into a single matrix, where each row corresponds to a time point, and each column represents a MEG channel. The function get_indices_from_list is used to generate indices marking where each session starts and ends within the concatenated data. Run the following commands to structure the data:

```
D_con = np.concatenate(D_raw, axis =0)
idx_data = statistics.get_indices_from_list(D_raw)
```

The generated indices for each session will look like this:

```
[[0 1530001]
[1530001 3034002]
...
[10496514 10864515]]
```

● **TIMING** <6 min

3. Load and prepare data: preprocessing data. In this example, we isolate brain activity in the alpha band (8–13 Hz) to focus on specific oscillatory patterns associated with attention and sensory suppression[37,38]. Preprocessing involves several steps:
   - Band-pass filtering. Apply a band-pass filter to extract the alpha frequency band (8–13 Hz). The same procedure can be applied to other frequency bands.
   - Standardization. Normalize the data to 0 mean.
   - Hilbert transform. Use the Hilbert transform to extract the amplitude (strength) and phase (timing) of brain waves.
   - PCA. Reduce data dimensionality by retaining 90% of the variance.
   - Downsampling. Reduce the sampling rate from 1,000 to 250 Hz to decrease computational load.
     Run the following code to preprocess the concatenated MEG data (D_con) and corresponding indices (idx_data).

```
# Define preprocessing parameters
freqs = (8, 13)# Alpha band
pca_variance = 0.9# Retain 90%
variance fs = 1000# Original sampling rate
f_target = 250# Target sampling rate after downsampling
standardise = True # Standardise the data
onpower = True # Hilbert transform
# Preprocess the data
D_preproc, idx_preproc = preproc.preprocess_data(
  data = D_con,
  indices= idx_data,
  fs=fs,
  standardise = standardise,
```

```
    filter= freqs,
    onpower= onpower,
    pca= pca_variance,
    downsample = f_target
  )
```

After preprocessing, the data are stored in `D_preproc`, and the corresponding indices are stored in `idx_preproc`.

● **TIMING <45 min**

4. Load and prepare data: set up and train an HMM. We use a standard Gaussian HMM to identify distinct brain states and track changes over time. The key output, `Gamma`, provides the probability of being in each state at every time point and forms the basis for subsequent analyses. The HMM is set up with the same parameters as in Procedure 1, but with five states:

```
K = 5
hmm_classic = glhmm.glhmm(model_beta ='no', K=K, covtype ='full')
```

Train the HMM by using the preprocessed MEG data (`data_session_preproc`) and session indices (`idx_data_preproc`):

```
Gamma, _, _= hmm_classic.train(
X=None,
Y= data_session_preproc,
indices= idx_data_preproc
)
```

The Gamma matrix contains the probability of each state at every time point.

● **TIMING <40 s**

5. Load and prepare data: configure HMM outputs for statistical analysis. With the Gamma values, we can analyze how brain states relate to specific events in the data. The Gamma matrix has the following dimensions: [2716140, 5]—downsampled from 10,864,515 to 2,716,140 time points, with 5 brain states. The reduction in time points reflects the downsampling from 1,000 to 250 Hz.
   • Epoch the data. To analyze responses to specific events (e.g., stimulus presentations), the Gamma data are divided into smaller segments called 'epochs'. Each epoch corresponds to a trial, defined by using event markers. The event marker time stamps need to be downsampled to match the Gamma data, by setting `fs_target` to 250 Hz. The window length for each epoch is set to 250 time points, representing a 1-s time window after the stimulus. Execute the following commands to extract the epochs:

```
fs_target= 250 # Define the target sampling frequency epoch_window_
tp = 250 # Epoch window length in timepoints
# Extract epochs for the HMM state time courses
gamma_epoch, idx_data_epoch, R_data_epoch = statistics.get_event_epochs(
D_data = Gamma,
R_data = R_data,
indices= idx_data_preproc,
event_markers= event_markers,
fs=fs,
fs_target=fs_target,
epoch_window_tp=epoch_window_tp
)
```

The resulting dimensions are:
- `gamma_epoch`: [250, 8368, 5]—250 time points per trial, 8,368 trials and 5 states
- `R_data_epoch`: [8368]—Stimulus labels for each trial (0 for inanimate, 1 for animate)
- `idx_data_epoch`: marks the start and end trial indices for each session
  This configuration prepares the `Gamma` data for statistical analyses by segmenting it into epochs aligned with the experimental events.

● **TIMING** <1 s

6. Statistical analysis: types of tests. We use the `test_across_trials` function from the statistics.py module to test whether the brain states (`gamma_epoch`, D) process the behavioral conditions (`R_data_epoch`, *R*)—watching animate versus inanimate objects—in the same way for each trial, or whether the responses vary across trials.

● **TIMING** <20 min

7. Statistical analysis: methods supported. Below, we show how to configure the required inputs and perform a multivariate statistical test with 10,000 permutations. For this analysis, we use brain data (`gamma_epoch`) and behavioral conditions (`R_data_epoch`) as inputs. Additional parameters include the indices for each session (`idx_data_epoch`), the number of permutations (`Nnull_samples`) and the analysis method (`method`). Run the following code:

```
# Set parameters for multivariate testing
method = "multivariate"
Nnull_samples = 10_000 # Number of permutations
# Perform across-trial testing
results_multivariate = statistics.test_across_trials(
  D_data = gamma_epoch,
  R_data = R_data_epoch,
  indices_blocks = idx_data_epoch,
  Nnull_samples= Nnull_samples,
  method = method
)
```

The results of the test are stored in the result_multivariate dictionary. For a breakdown of the dictionary structure, see Box 1 in Procedure 1.
▲ **CRITICAL** The option `test_statistics_option=True` is required only if cluster-level inference will be used during multiple testing correction at Step 8. By default, it is set to True.

● **TIMING** <5 s

8. Statistical analysis: multiple testing correction and cluster statistics. In this example, we demonstrate how to perform cluster-level inference, which identifies clusters of significant results while reducing the risk of false positives. The correction uses the output from result_multivariate, and the test focuses on *P* values below a threshold of 0.01, specified by `alpha = 0.01`. The `pval_cluster_based_correction` function performs the correction:

```
pval_cluster = statistics.pval_cluster_based_correction(
results_multivariate,
alpha = 0.01
)
```

The function returns a pval cluster, which contains the adjusted *P* values after cluster-level inference.
▲ **CRITICAL** Alternative multiple testing correction procedures, such as the MaxT method (FWER correction), Bonferroni or Benjamini–Hochberg, can also be used.

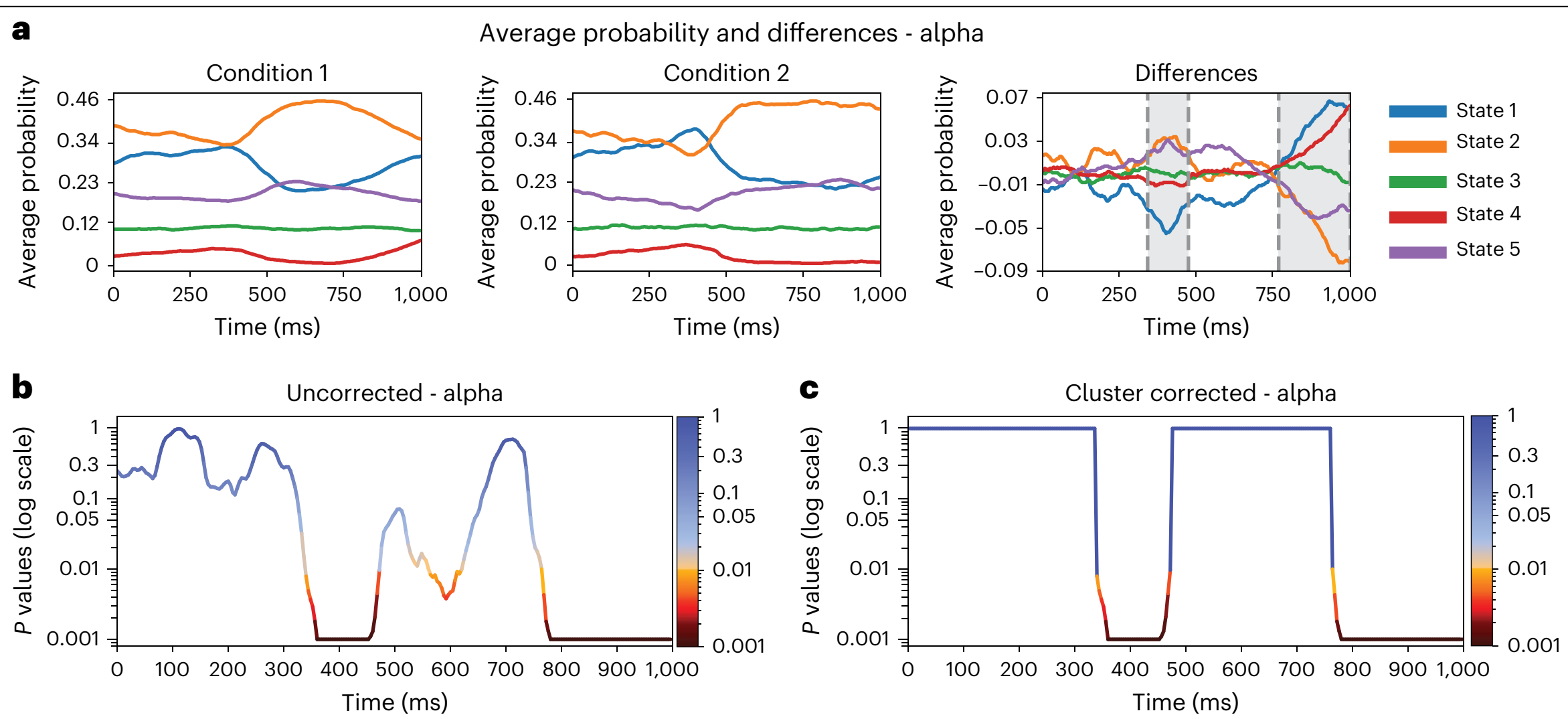

**Fig. 4 | Result from Procedure 2. a**, Average state probabilities over time for inanimate and animate stimuli and the difference between the two conditions. Significant differences after cluster-level inference are highlighted in gray. **b**, Results of the multivariate test (uncorrected). **c**, Results of the multivariate test after applying cluster-level inference.

● **TIMING** <5 s

9. Visualization: visualizing statistical results. This step visualizes the results for both uncorrected and cluster-corrected *P* values from the multivariate test, as shown in Fig. 4. The function `plot_p_values_over_time` from the graphics module is used to generate line plots. For this example, `alpha = 0.01` is set to highlight *P* values below this threshold.

```
# Set parameters
xlabel = "Time (ms)"
alpha = 0.01
# Plot uncorrected p-values
graphics. plot_p_values_over_time(
  results_multivariate ["pval"],
  title_text=f"Uncorrected - Alpha",
  xlabel= xlabel,
  alpha = alpha,
  )
# Plot cluster corrected p-values
graphics.plot_p_values_over_time(
  pval_cluster,
  title_text=f"Cluster Corrected - Alpha",
  xlabel= xlabel,
  alpha = alpha,
)
```

## Procedure 3: across-sessions-within-subject testing

▲ **CRITICAL** Before starting this procedure, follow Steps 1–5 from Procedure 2 for data and preprocessing setup. This procedure then focuses on the statistical analysis for the across-sessions-within-subject test.

● **TIMING** <1 s

1. Load the data into the Python environment, as in Step 1 of Procedure 2.
2. Perform data structuring for the HMM, as in Step 2 of Procedure 2.
3. Preprocess the data, as in Step 3 of Procedure 2.
4. Set up and train the HMM, as in Step 4 of Procedure 2.
5. Configure HMM outputs for statistical analysis, as in Step 5 of Procedure 2.
6. Statistical analysis: types of tests. We use the `test_across_sessions_within_subjects` function from the `statistics.py` module to test whether the brain states (`gamma_epoch`, *D*) encode the behavioral conditions (`R_data_epoch`, *R*)—watching animate versus inanimate objects—consistently across sessions or if the encoding changes. Differences may suggest that the brain processes the same task differently across different sessions over time.

● **TIMING** <17 min

7. Statistical analysis: methods supported. Below, we show how to configure the required inputs and perform a multivariate statistical test with 10,000 permutations. For this analysis, we use brain data (`gamma_epoch`) and behavioral conditions (`R_data`) as inputs. Additional parameters include the indices for each session (`idx_data_epoch`), the number of permutations (`Nnull_samples`) and the analysis method (`method`). Run the following code:

```
# Set parameters for multivariate testing
method = "multivariate"
Nnull_samples = 10_000 # Number of permutations
# Perform across-trials testing
results_multivariate = statistics.test_across_sessions_within_subject(
  D_data = gamma_epoch,
  R_data = R_data_epoch,
  indices_blocks = idx_data_epoch,
  Nnull_samples= Nnull_samples,
  method = method
)
```

The results of the test are stored in the `result_multivariate` dictionary. For a breakdown of the dictionary structure, see Box 1 in Procedure 1.

● **TIMING** <5 s

8. Statistical analysis: multiple testing correction and cluster statistics. In this example, we demonstrate how to apply multiple testing correction by using FWER correction with the MaxT method. To run the test, we use the function `pval_FWER_correction`, and it requires only the `result_multivariate` dictionary as input.

```
pval_FWER = statistics.pval_FWER_correction(result_multivariate)
```

The function returns the FWER-corrected *P* values in the variable pval_FWER.

● **TIMING** <5 s

9. Visualization: visualizing statistical results The results for uncorrected, FWER-corrected, Benjamini–Hochberg and cluster-corrected *P* values are shown in Fig. 5 for both multivariate and univariate tests. Although the code example and text focus on the multivariate test with FWER correction, the figure provides a broader overview of different correction methods. Notably, because this dataset includes only one variable (stimulus presentation of animate and inanimate objects), FWER correction has no effect on the multivariate test results. This is expected, because MaxT correction applies only when multiple tests are performed. However, in the univariate test, FWER correction does show an effect, because the permutation process involves multiple tests across time points.
To perform a univariate test, set `method="univariate"` in the statistical testing function.

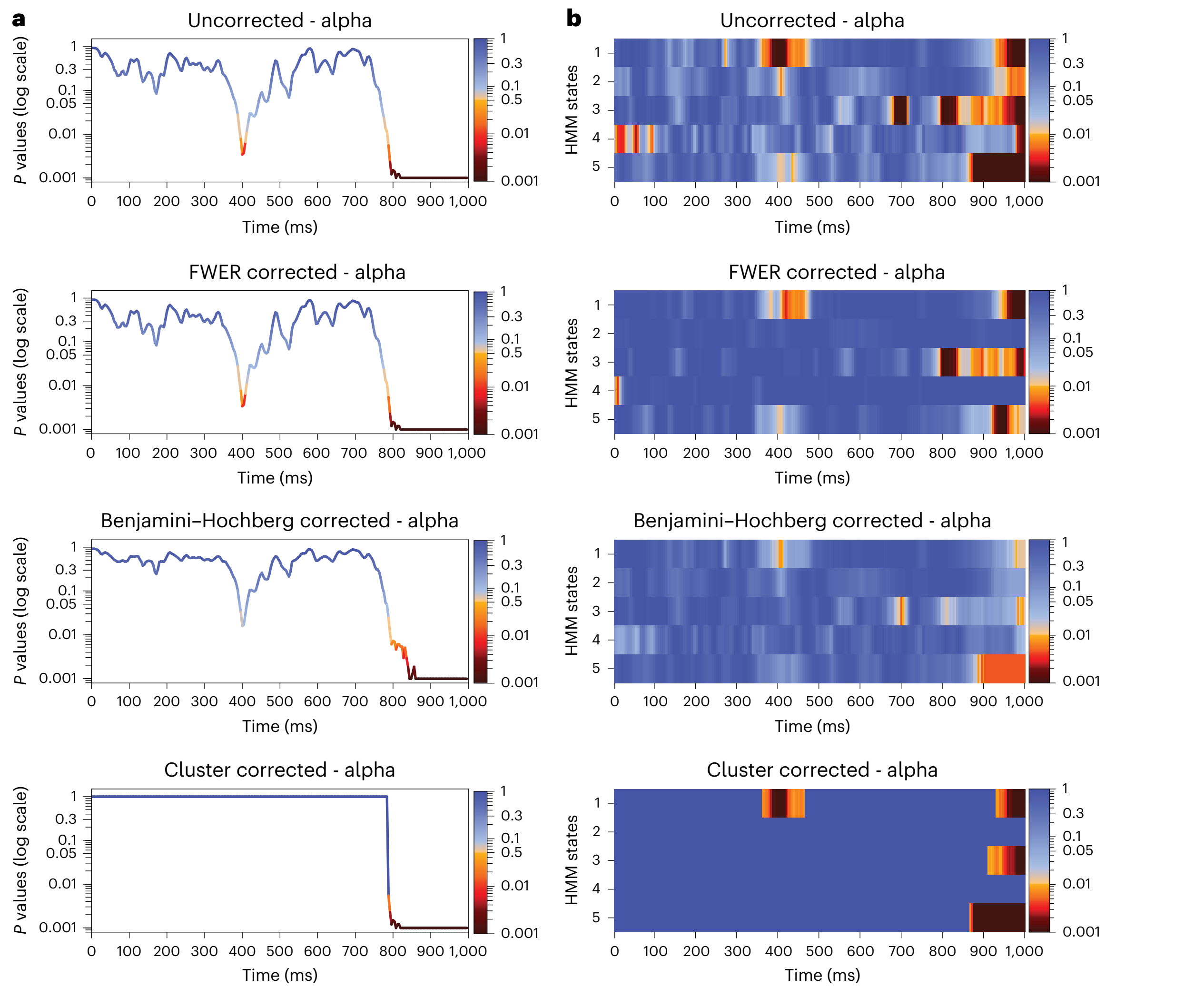

**Fig. 5 | Result from Procedure 3. a**, Multivariate test results for uncorrected, FWER-corrected and Benjamini–Hochberg corrected *P* values and cluster-level inference. **b**, Univariate test results for the same correction methods.

Line plots visualize multivariate tests, and heatmaps are used for univariate tests. Both methods are part of the graphics module. For this example, `alpha = 0.01` is set to highlight *P* values below the threshold.

```
# Threshold for the p-value plots
alpha = 0.01
# Plot uncorrected p-values graphics.
plot_p_values_over_time(
  results_multivariate[" pval"],
  title_text=f"incorrected - Alpha",
  xlabel= xlabel,
  alpha = alpha,
)
# Plot FWER corrected p-values
graphics.plot_p_values_over_time(
  pval_FWER,
  title_text=f"FWER - Alpha",
```

```
  xlabel= xlabel,
  alpha = alpha
)
```

## Procedure 4: across-state-visits testing

● **TIMING** <5 s

1. Load and prepare data: load data into the Python environment. For Procedure 4, we analyze data collected from 10 participants during resting-state MEG recordings. Each participant completed two sessions, except for one participant who completed only one session. The MEG data, stored in `data_meg.pkl`, is a list in which each element corresponds to a session. It is loaded into the variable `data_meg`, referred to as 'matrix *D*'. Pupillometry, recorded simultaneously for each session, is stored in `pupillometry.pkl` and loaded into the variable `data_pupillometry`, referred to as 'matrix *R*'. Unlike in previous protocols, in which the HMM was trained on the dataset, we use a pre-trained temporal delayed embedding HMM (TDE-HMM) from ref. 15. We use this pre-trained model to decode brain states for each session of MEG data without requiring additional training. The pre-trained model is stored in the MATLAB file `hmm.mat`. To decode the MEG data, load the pre-trained TDE-HMM using the `read_flattened_hmm_mat` function from the `io` module.
   - Retrieve the model settings by using `scipy.io.loadmat`. We use the pickle module to load the MEG and pupillometry data:

   ```
   with open("data_meg.pkl", "rb") as f:
     data_meg = pickle.load(f)
   with open("pupillometry.pkl", "rb") as f:
     data_pupillometry = pickle.load(f)
   # Load pre-trained TDE - HMM
   hmm_TDE = io.read_flattened_hmm_mat("hmm.mat")
   # Load the settings of the TDE - HMM
   hmm_TDE_settings = scipy.io.loadmat('hmm.mat')
   ```

   In summary:
   - `data_meg`: brain activity as a list of 19 sessions, where each session is a 2D matrix with shape [No. of time points × No. of channels] (42 channels extracted by using PCA)
   - `data_pupillometry`: pupil size as a list of 19 sessions, where each session is a 1D array with shape [No. of time points]
   - `hmm_TDE`: pre-trained TDE-HMM model[15] used to decode brain states from brain data
     ▲ **CRITICAL** Ensure that the MEG and pupillometry data are temporally aligned for accurate state decoding and statistical testing. The number of time points in data_meg and data_pupillometry must match for each session. Any mismatched data lengths could lead to errors during analysis.

● **TIMING** <5 s

2. Load and prepare data: data structuring for the HMM. Before applying the pre-trained TDE-HMM model, we need to organize the `data_meg` to a 2D matrix with the shape [(No. of time points across all sessions) × No. of channels]. This involves concatenating the data from all 19 sessions into a single matrix, where each row corresponds to a time point, and each column represents a MEG channel. The pupil size data (`data_pupillometry`) also needs to be structured along the time dimension to form a single 1D array: [No. of time points across all sessions]. To track the start and end time points for each session, we generate an index matrix by using the function `get_indices_from_list`. The resulting matrix has the shape [No. of sessions × 2], where each row specifies the start and end time points for a session. Run the following commands to structure the data:

```
D_con = np.concatenate(data_meg, axis =0)
R_data = np.concatenate(data_pupillometry, axis =0)
idx_data = statistics.get_indices_from_list(data_meg)
```

The generated indices for each session will look like this:

```
[[0 85996]
[85996 167903]
…
[10496514 1473460]]
```

● **TIMING** <20 s

3. Load and prepare data: preprocessing data. Before analyzing the data by using TDE-HMM, the MEG data must be formatted correctly. This preparation involves two main steps:
   - Preprocessing the brain data (`D_preproc`). We standardize the data to ensure that all time series data are on the same scale to ensure comparability. This step uses the `preprocess_data` function to standardize the data to a 0 mean and unit variance.
   - Preparing data for the TDE-HMM (D_tde). The `build_data_tde` function prepares the MEG data for TDE-HMM analysis by (i) adding time lags to capture changes in brain activity over short windows (for this example, seven time lags before and after each time point are used) and (ii) applying a PCA projection to reduce dimensionality, with settings extracted from `hmm_TDE_settings`.

   We use the following script to preprocess the data:

```
# Preprocess data
D_preproc, idx_data_preproc = preproc.preprocess_data(
data = D_con,
indices= idx_data,
standardise =True, # Standardise the data
)
# Specify time lags embedded_lags = 7
lags = np.arange(-embedded_lags, embedded_lags + 1)
# Load PCA projection settings
pca_proj = hmm_TDE_settings["train"][" A"][0][0]
# Build the MEG data in TDE format
D_tde, indices_tde = preproc.build_data_tde(
data = D_preproc,
indices =idx_data_preproc, lags=lags,
pca= pca_proj
)
```

   Now the MEG data are ready for decoding brain states with the pre-trained TDE-HMM model.

● **TIMING** <1 min

4. Load and prepare data: set up and train an HMM. We can use the TDE-HMM to decode brain activity into distinct states over time. These states form a sequence called the 'Viterbi path' (`D_vpath_tde`), which shows the brain's most likely state at each time point. The across-state-visits test is the only statistical test in this framework that relies on the Viterbi path instead of other outputs, such as state time courses (Gamma). We can decode the Viterbi path by using the following command:

```
D_vpath_tde = hmm_TDE.decode(X=None, Y= D_tde, indices= indices_tde,
viterbi= True)
```

▲ **CRITICAL** Ensure that the Viterbi path (`D_vpath_tde`) is decoded correctly, because it is the primary input for the across-state-visits test. Errors in decoding or preprocessing may lead to misleading conclusions, so always verify the input data format and preprocessing steps before decoding.

● **TIMING** <40 s

5. Load and prepare data: configure HMM outputs for statistical analysis. To analyze the relationship between brain states from the Viterbi path (`D_vpath_tde`, *D*) and pupil size (`R_data`, *R*), the datasets must be aligned. The dimensions of the data are as follows:
   - Viterbi path (`D_vpath_tde`): (1473194,12)
   - Pupil size (`R_data`): (1473460,)

   The difference in length occurs because the TDE-HMM introduces a lag of 7 time points at the start and end of each session and thereby removes 14 time points per session. To match the dimensions, the Viterbi path is padded to restore the original length of the pupillometry data. This is achieved by using the `pad_vpath` function, which adjusts for the lagged time points on the basis of session boundaries.

```
embedded_lags = 7
D_vpath_pad = statistics.pad_vpath(
vapth = D_vpath_tde,
lag_val = embedded_lags,
indices_tde = indices_tde
)
```

   The padded Viterbi path (`D_vpath_pad`) is initially stored as a 2D array with one-hot encoding, where each row represents a time point, and one state is active per row. To simplify the data and reduce memory usage, we convert the array to a 1D format where each value represents the active state for a given time point and store the data into the variable `D_vpath`.

```
D_vpath = D_vpath_pad.nonzero()[1] + 1
```

   ▲ **CRITICAL** Alignment of the Viterbi path and pupil size data is essential for performing the statistical testing. Always verify that both datasets match in length after padding. In addition, plotting FO can provide a useful overview of how consistently the TDE-HMM captures brain activity across sessions, as shown in Fig. 6a.

● **TIMING** <1 s

6. Statistical analysis: types of tests. We use the `test_across_state_visits` function from the `statistics.py` module to test whether specific brain states (`D_vpath`, *D*) are associated with differences in pupil size (`R_data`, *R*) during resting-state recordings.

● **TIMING** 3–4 h

7. Statistical analysis: methods supported. Across-state-visits analysis includes methods such as OSA and OSR to explore how brain states relate to other signals, like pupil size in our case. Here, we focus on OSA. The OSA test compares pupil size between pairs of brain states. For example, it tests whether the average pupil size during state 1 differs from that in state 2, state 3 and so on. This helps reveal how specific brain states influence pupil size during resting-state recordings. To perform this analysis, we use the brain state sequence (`D_vpath`) and pupil size data (`R_data`) as inputs. Additional settings include the number of permutations (`Nnull_samples`) and the test type (`method`). Run the following code to perform the test:

```
# Set parameters for the state pair comparison test
method = "OSA"
```

```
Nnull_samples = 10_000 # Number of Viterbi path surrogates
# Run the analysis
results_OSA = statistics.test_across_state_visits(
  D_data = D_vpath,
  R_data = R_data,
  method = method,
Nnull_samples= Nnull_samples
)
```

The test results are stored in a variable called '`results_OSA`'. For a breakdown of the dictionary structure, see Box 2.

▲ **CRITICAL** Creating the permutation matrix is the most time-intensive part of the test because it involves every time point in the data. For this example, the matrix has a size of [1473460, 1000] [No. of time points, No. of null samples]. To save time, you can create this matrix ahead of time and store it (e.g., as `vpath_surrogates`). Using this precomputed matrix reduces the test run time to just a couple of minutes.

● **TIMING** <5 s

8. Statistical analysis: multiple testing correction and cluster statistics. In this example, we apply multiple testing correction using the Benjamini–Hochberg procedure. The function `pval_correction` performs this correction and requires the *P* values from `results_OSA` as input, with the method set to 'fdr_bh'.

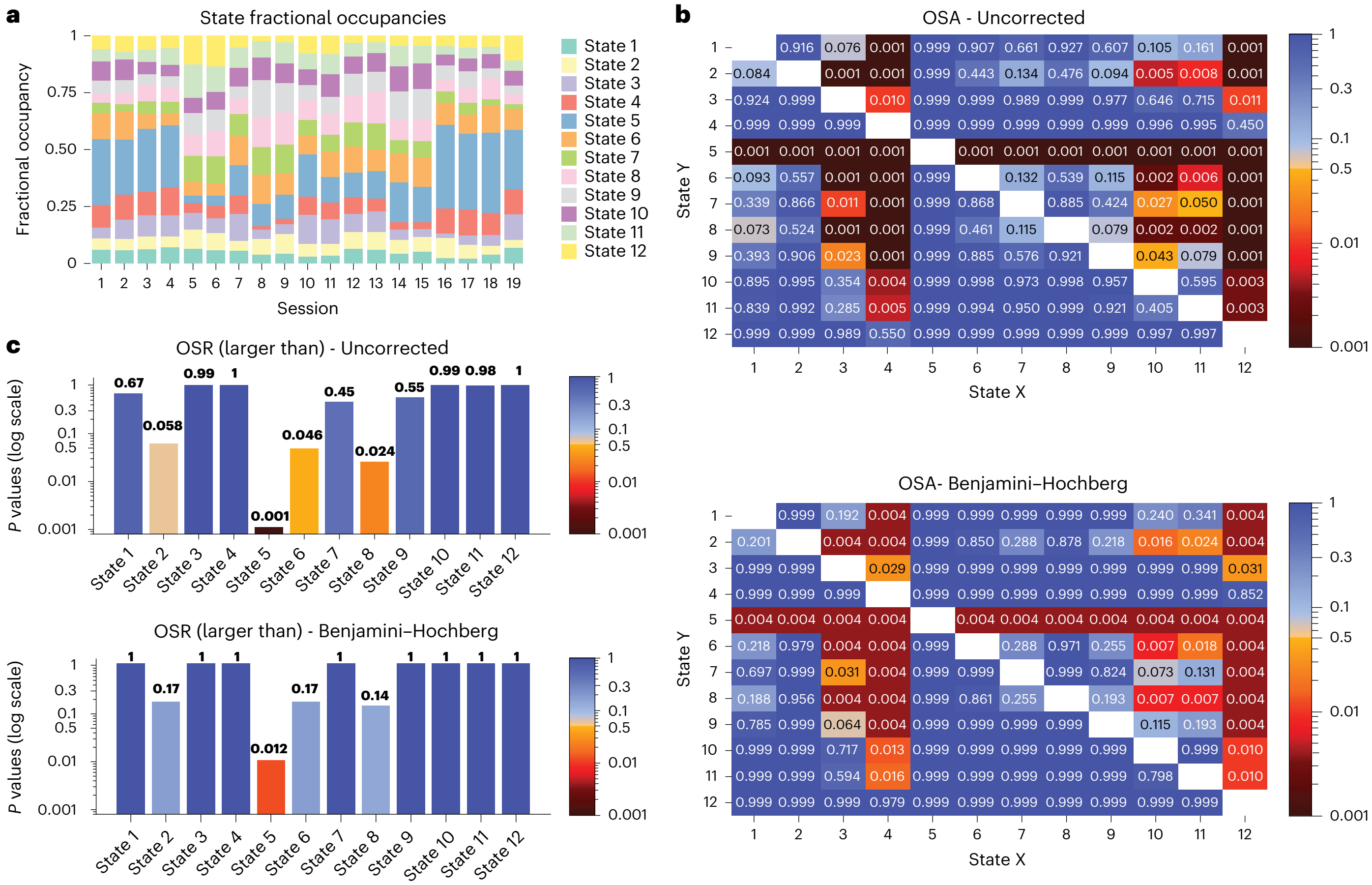

**Fig. 6 | Results from Procedure 4. a**, FO for each session, showing the stability of TDE-HMM decoding. **b**, OSA test results, with both uncorrected and Benjamini–Hochberg-corrected *P* values. **c**, OSR test results, shown for comparison, with the same correction methods applied.

## BOX 2

## Details of the results dictionary

- 'pval': *P* values computed under the null hypothesis.
- 'base_statistics': the observed test statistic calculated from the original (unshuffled) Viterbi path.
- 'null_stat_distribution': test statistics generated under the null hypothesis, where the first row corresponds to the observed test statistic ('base_statistics').
- 'statistical_measures': dictionary specifying the type of test statistic in each column in ('base_statistics'), such as *t*-statistics or *F*-statistics.
- 'test_type': type of test performed (across_state_visits).
- 'method': analytical approach used (e.g., multivariate and univariate).
- 'max_correction': whether Max-statistic correction was used for multiple comparisons.
- 'Nnull_samples': total number of Monte Carlo samples (i.e., surrogate Viterbi paths) including the observed one.
- 'test_summary': dictionary summarizing the test results.
- 'pval_f_multivariate': *F* test *P* values for multivariate tests and Nnull_samples > 1.
- 'pval_t_multivariate': *t* test *P* values for multivariate tests and Nnull_samples > 1.

```
# Apply Benjamini - Hochberg correction
pval_fdr_bh, _ = statistics.pval_correction(
  results_OSA,
  method ='fdr_bh'
)
```

The corrected *P* values are stored in the variable pval_fdr_bh.

● **TIMING** <5 s

9. Visualization: visualizing statistical results. For Procedure 4, we performed only the OSA test, which compares pupil size between pairs of brain states. However, we also visualize OSR results to provide a reference for both methods. Figure 6 displays the uncorrected and Benjamini–Hochberg–corrected *P* values for OSA and OSR. For OSA, *P* values are stored in a $[p, p]$ array where $p = 12$ (representing 12 states). Values above the diagonal represent comparisons where state $X >$ state $Y$, whereas values below the diagonal represent state $X <$ state $Y$. To run an OSR test, set `method="OSR"` in the statistical testing function. Heatmaps are used for OSA results, while bar plots are used for OSR. Both visualization functions are part of the graphics module.

```
# Plot uncorrected p-values
graphics.plot_p_value_matrix(
  results_OSA ["pval"],
  title_text ='OSA - Uncorrected',
  xlabel="State X",
  ylabel="State Y",
  alpha =0.05,
  none_diagonal=True,
  annot=True,
  x_tick_min =1,
  x_tick_max =12
)
# Plot Benjamini - Hochberg corrected p- values
graphics. plot_p_value_matrix(
  pval_fdr_bh,
  title_text ='OSA - Benjamini - Hochberg correction',
```

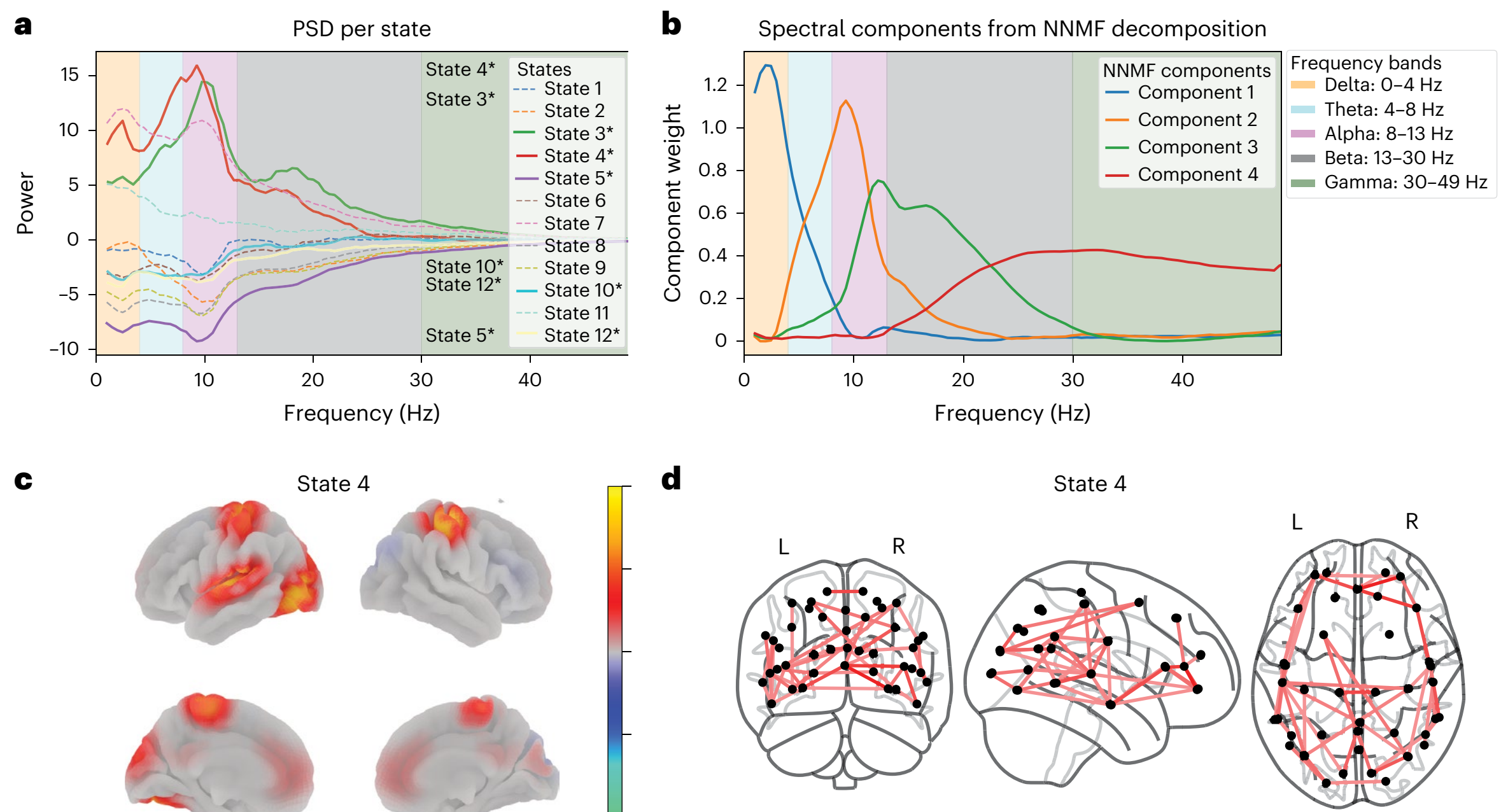

**Fig. 7 | Spectral and spatial characterization of brain states. a**, Power spectral density (PSD) profiles for each brain state inferred from the HMM. Highlighted lines indicate states identified as significant by the univariate test. **b**, Spectral components obtained by applying NNMF with four components to the PSDs. **c**, Brain map showing the first NNMF component for state 4, showing higher activation in sensory (visual, somatosensory and auditory) areas. **d**, Connectivity map for state 4 based on the first NNMF component. L, left; R, right.

```
    figsize = (9, 5),
    xlabel= "State X",
    ylabel= "State Y",
    alpha =0.05,
    none_diagonal=True,
    annot=True,
    x_tick_min =1,
    x_tick_max =12
)
```

In addition to statistical testing, it is useful to explore the spectral and spatial characteristics of the decoded brain states. These features help to describe the functional profile of each state and support interpretation of the results. Figure 7 shows various visualizations including power spectra, spectral components from a data-driven decomposition (non-negative matrix factorization or non-negative matrix factorization (NNMF)[15]) and spatial maps of power and coherence for a single spectral component. All visualization steps are implemented in the Procedure 4 notebook.

## Timing

Running the full protocol, from preprocessing through statistical testing and visualization, can be completed in about 2–5 h in the example datasets shown here. Preprocessing (Steps 1–3) typically requires about 5–15 min, set-up and training of the HMM (Steps 3–5) takes about 1–4 h and statistical testing with result visualization (Steps 6–9) takes about 2 min to 4 h, depending on chosen settings. These estimates are based on a Lenovo ThinkPad T16 Gen 3 laptop (Intel Core Ultra 7 155U, 32 GB of random access memory (RAM), 1 TB solid state drive (SSD)). Of course, the processing time depends on the dataset size, type of test and applied method; thus, actual run times may vary.

## Anticipated results

The GLHMM framework offers an accessible yet effective set of tools for analyzing temporal dynamics that could be used across different fields of research, although we have focused here on neuroscience applications. Using the four statistical tests presented, users can investigate associations between the properties of a dynamic system and a set of external variables. For instance, the across-subjects test assesses associations between brain states and individual traits or characteristics, while the across-trials test can pinpoint temporal patterns in experimental conditions. For longitudinal studies' benefit, the across-sessions-within-subject test can assess changes in brain–behavior relationships over longer time scales. Finally, the across-state-visits test can be used to probe the interactions between brain states and concurrently recorded signals.

To illustrate the types of results that can be obtained, all raw data are available on Zenodo (https://doi.org/10.5281/zenodo.15213970). The full analysis, including intermediate outputs, can be reproduced directly by using the Jupyter notebooks provided in the associated GitHub repository (https://github.com/Nick7900/glhmm_protocols), which download the data from scratch and guide users through each step of the workflow.

## Ethics declarations

Procedure 1 used HCP data (ethics approval obtained by the HCP consortium), and Procedures 2–4 used anonymized pilot datasets collected on members of our research group, for which no additional ethical approval was required.

### Reporting summary

Further information on research design is available in the Nature Portfolio Reporting Summary linked to this article.

### Data availability

All data used in this protocol are freely accessible on Zenodo (https://doi.org/10.5281/zenodo.15213970). The repository also includes a link to the associated code on GitHub: https://github.com/Nick7900/glhmm_protocols.

### Code availability

All code is available on GitHub (https://github.com/Nick7900/glhmm_protocols), provided as Python notebooks that can be run directly in the cloud by using Google Colab, so there is no need to install Python or any packages locally. For reproducibility, the repository is also archived on Zenodo at https://zenodo.org/records/15213970 (https://doi.org/10.5281/zenodo.15213970). This setup supports versioning for future updates, including new code and tutorials.

## Acknowledgements

We sincerely thank F. Fardo for collecting the MEG data used in Procedures 2 and 3. D.V. is supported by a Novo Nordisk Foundation Emerging Investigator Fellowship (NNF19OC-0054895) and an ERC Starting Grant (ERC-StG-2019-850404).

## Author contributions

N.Y.L. provided conceptualization, formal analysis, software, writing of the original draft, review and editing of the original draft and visualization. L.B.P. provided data preparation, review and editing of the original draft and validation. C.A. provided conceptualization, methodology, review and editing of the original draft. A.M.W. provided conceptualization, methodology, review and editing of the original draft. D.V. provided conceptualization, methodology, software, writing of the original draft, review and editing of the original draft and funding acquisition.

## Competing interests

The authors declare no competing interests.

## Additional information

**Correspondence and requests for materials** should be addressed to Nick Y. Larsen.

**Peer review information** *Nature Protocols* thanks Aiping Liu and the other, anonymous, reviewer(s) for their contribution to the peer review of this work.

# nature portfolio

Corresponding author(s): NP-P250303B

Last updated by author(s): Sep 17, 2025

# Reporting Summary

Nature Portfolio wishes to improve the reproducibility of the work that we publish. This form provides structure for consistency and transparency in reporting. For further information on Nature Portfolio policies, see our Editorial Policies and the Editorial Policy Checklist.

## Statistics

For all statistical analyses, confirm that the following items are present in the figure legend, table legend, main text, or Methods section.

| n/a | Confirmed | |
|---|---|---|
| ☐ | ☒ | The exact sample size (*n*) for each experimental group/condition, given as a discrete number and unit of measurement |
| ☐ | ☒ | A statement on whether measurements were taken from distinct samples or whether the same sample was measured repeatedly |
| ☐ | ☒ | The statistical test(s) used AND whether they are one- or two-sided<br>*Only common tests should be described solely by name; describe more complex techniques in the Methods section.* |
| ☐ | ☒ | A description of all covariates tested |
| ☐ | ☒ | A description of any assumptions or corrections, such as tests of normality and adjustment for multiple comparisons |
| ☐ | ☒ | A full description of the statistical parameters including central tendency (e.g. means) or other basic estimates (e.g. regression coefficient) AND variation (e.g. standard deviation) or associated estimates of uncertainty (e.g. confidence intervals) |
| ☐ | ☒ | For null hypothesis testing, the test statistic (e.g. *F*, *t*, *r*) with confidence intervals, effect sizes, degrees of freedom and *P* value noted<br>*Give P values as exact values whenever suitable.* |
| ☒ | ☐ | For Bayesian analysis, information on the choice of priors and Markov chain Monte Carlo settings |
| ☒ | ☐ | For hierarchical and complex designs, identification of the appropriate level for tests and full reporting of outcomes |
| ☐ | ☒ | Estimates of effect sizes (e.g. Cohen's *d*, Pearson's *r*), indicating how they were calculated |

*Our web collection on statistics for biologists contains articles on many of the points above.*

## Software and code

Policy information about availability of computer code

| | |
|---|---|
| Data collection | Procedure 1 used the Human Connectome Project (HCP) dataset, which was collected and distributed by the HCP consortium. Procedures 2–4 used datasets collected on members of our research group using an Elekta MEG system. |
| Data analysis | All analyses were performed in Python (version 3.10). Custom analyses were conducted using the open-source GLHMM package (available at https://github.com/vidaurre/glhmm), together with standard scientific libraries including NumPy (v1.23.5), SciPy (v1.10.1), and scikit-learn (v1.2.1). Code for reproducing the results is publicly available here (https://github.com/Nick7900/glhmm_protocols). |

For manuscripts utilizing custom algorithms or software that are central to the research but not yet described in published literature, software must be made available to editors and reviewers. We strongly encourage code deposition in a community repository (e.g. GitHub). See the Nature Portfolio guidelines for submitting code & software for further information.

## Data

Policy information about availability of data

All manuscripts must include a data availability statement. This statement should provide the following information, where applicable:

- Accession codes, unique identifiers, or web links for publicly available datasets
- A description of any restrictions on data availability
- For clinical datasets or third party data, please ensure that the statement adheres to our policy

All data used in this protocol are freely accessible on Zenodo (DOI: 10.5281/zenodo.15213970). The repository also includes a link to the associated code and tutorials on GitHub: https://github. com/Nick7900/glhmm_protocols.

## Research involving human participants, their data, or biological material

Policy information about studies with human participants or human data. See also policy information about sex, gender (identity/presentation), and sexual orientation and race, ethnicity and racism.

| | |
|---|---|
| Reporting on sex and gender | For Procedure 1 (HCP dataset), sex and gender information is available and was collected under the HCP consortium's protocols. For Procedures 2–3 (datasets collected internally on members of our research group), each involved a single female participant. For Procedure 4, 8/10 participants were male. The sex distribution in Procedures 2–4 is not balanced, as these pilot datasets were collected for methodological development rather than for drawing biological conclusions. |
| Reporting on race, ethnicity, or other socially relevant groupings | For Procedure 1 (HCP dataset), race and ethnicity information is available through the HCP consortium. For Procedures 2–4, all participants were white Caucasian, reflecting the composition of our research group at the time of data collection. |
| Population characteristics | Procedure 1 used the HCP dataset, which includes demographic information as described by the HCP consortium. Procedure 2 and 3 involved a single pilot participant from our research group. Procedure 4 involved ten participants from our research group. No demographic or clinical information was collected for Procedures 2–4. |
| Recruitment | The Human Connectome Project dataset was recruited and made available by the HCP consortium. The additional datasets were collected on members of our research group for methodological development purposes. |
| Ethics oversight | The HCP dataset was acquired under the HCP consortium's ethical approvals (Procedure 1). The pilot datasets collected on members of our research group (Procedures 2–4) did not include personal identifying information and therefore did not require additional ethical approval under institutional policy. |

Note that full information on the approval of the study protocol must also be provided in the manuscript.

# Field-specific reporting

Please select the one below that is the best fit for your research. If you are not sure, read the appropriate sections before making your selection.

☒ Life sciences ☐ Behavioural & social sciences ☐ Ecological, evolutionary & environmental sciences

For a reference copy of the document with all sections, see nature.com/documents/nr-reporting-summary-flat.pdf

# Life sciences study design

All studies must disclose on these points even when the disclosure is negative.

| | |
|---|---|
| Sample size | No formal sample size calculation was performed. The datasets were selected to provide proof-of-concept demonstrations of the framework across different contexts. Procedure 1 used the HCP dataset (a large publicly available sample), Procedures 2–3 each used one participant, and Procedure 4 used ten participants from our research group. These datasets were collected as pilot data for methodological development rather than for drawing biological conclusions. The sample sizes were sufficient for the statistical framework, which relies on permutation testing and Monte Carlo resampling rather than on parametric assumptions about sample size. |
| Data exclusions | One session was excluded in Procedure 4 because of malfunctioning of the equipment. No other data were excluded beyond standard preprocessing steps (e.g. artefact rejection). |
| Replication | The analyses can be fully replicated using the notebooks and code provided with the paper. Replication of biological findings was not the aim; the focus was on reproducibility of the framework's procedures across independent datasets. |
| Randomization | Randomisation of participants was not relevant, as no experimental groups were defined. |
| Blinding | Blinding was not relevant to this study. The datasets were analysed for methodological development and proof-of-concept demonstrations rather than for testing biological or clinical hypotheses. |

# Reporting for specific materials, systems and methods

We require information from authors about some types of materials, experimental systems and methods used in many studies. Here, indicate whether each material, system or method listed is relevant to your study. If you are not sure if a list item applies to your research, read the appropriate section before selecting a response.

## Materials & experimental systems

| n/a | Involved in the study |
|---|---|
| ☒ | ☐ Antibodies |
| ☒ | ☐ Eukaryotic cell lines |
| ☒ | ☐ Palaeontology and archaeology |
| ☒ | ☐ Animals and other organisms |
| ☒ | ☐ Clinical data |
| ☒ | ☐ Dual use research of concern |
| ☒ | ☐ Plants |

## Methods

| n/a | Involved in the study |
|---|---|
| ☒ | ☐ ChIP-seq |
| ☒ | ☐ Flow cytometry |
| ☒ | ☐ MRI-based neuroimaging |

## Plants

Seed stocks: n/a

Novel plant genotypes: n/a

Authentication: n/a